\documentclass[preprint]{ptephy}

\preprintnumber{KYUSHU-HET-226}
\usepackage{amssymb}
\usepackage{amsthm}
\usepackage{amsmath}
\usepackage{booktabs}
\usepackage{bbm}
\usepackage{bm}
\usepackage{mathtools}
\usepackage{simplewick}
\usepackage{subcaption}

\allowdisplaybreaks

\DeclareMathOperator{\tr}{tr}

\newcommand{\Slash}[1]{{\ooalign{\hfil/\hfil\crcr$#1$}}}
\numberwithin{equation}{section}

\begin{document}

\title{Gradient flow exact renormalization group---inclusion of fermion
fields---}

\author{%
\name{\fname{Yuki} \surname{Miyakawa}}{1} and
\name{\fname{Hiroshi} \surname{Suzuki}}{1,\ast}
}

\address{%
\affil{1}{Department of Physics, Kyushu University, 744 Motooka, Nishi-ku,
Fukuoka 819-0395, Japan}
\email{hsuzuki@phys.kyushu-u.ac.jp}
}

\date{\today}

\begin{abstract}
The gradient flow exact renormalization group (GFERG) is an exact
renormalization group motivated by the Yang--Mills gradient flow and its
salient feature is a manifest gauge invariance. We generalize this GFERG,
originally formulated for the pure Yang--Mills theory, to vector-like gauge
theories containing fermion fields, keeping the manifest gauge invariance. For
the chiral symmetry we have two options: one possible formulation preserves
the conventional form of the chiral symmetry and the other simpler formulation
realizes the chiral symmetry in a modified form \`a la Ginsparg--Wilson. We
work out a gauge-invariant local Wilson action in quantum electrodynamics to
the lowest nontrivial order of perturbation theory. This Wilson action
reproduces the correct axial anomaly in~$D=2$.
\end{abstract}

\subjectindex{B05, B32}
\maketitle

\section{Introduction}
\label{sec:1}
The Wilson exact renormalization group (ERG)~\cite{Wilson:1973jj} (see also
Refs.~\cite{Wegner:1972my,Wegner:1972ih}; for reviews, see
Refs.~\cite{Morris:1993qb,Pawlowski:2005xe,Igarashi:2009tj,Rosten:2010vm,%
Dupuis:2020fhh}) is important because, among many other things, it provides a
unique framework to consider possible quantum field theories beyond
perturbation theory. Given specific field contents, all possible quantum field
theories are obtained by the continuum limit around each fixed point of the
ERG equation with that field contents. In this sense, we may regard ERG as a
theory of theories.

In particle physics, gauge symmetry is a fundamental principle and we are thus
interested in ERG trajectories, i.e.\ solutions of the ERG equation, which
preserve this symmetry. Traditional formulations, however, employ the momentum
cutoff to define the ERG transformation, and this cutoff explicitly breaks the
gauge symmetry in the conventional form. Although it is possible to define a
modified gauge transformation which is consistent with the ERG
evolution~\cite{Becchi:1996an} (see also
Refs.~\cite{Igarashi:2019gkm,Igarashi:2009tj} and references cited therein),
and in principle one can maintain the gauge invariance in the modified form,
since such a transformation depends on the Wilson action itself, it appears
very hard to determine a nonperturbative truncation of the Wilson action being
consistent with this exact symmetry of ERG. For nonperturbative applications of
ERG in particle physics, therefore, a manifestly gauge-invariant ERG
formulation is highly desirable. Such formulations have been developed, for
instance in~Refs.~\cite{Morris:1998kz,Morris:1999px,Morris:2000fs,%
Arnone:2005fb,Morris:2006in,Wetterich:2016ewc,Wetterich:2017aoy}.

The gradient flow exact renormalization group (GFERG) proposed
in~Ref.~\cite{Sonoda:2020vut} is one such manifestly gauge-invariant ERG
formulation. This formulation is motivated by a similarity between the course
graining process in ERG and the diffusion of a field configuration in
spacetime. In particular, the diffusion defined by the Yang--Mills gradient
flow~\cite{Narayanan:2006rf,Luscher:2009eq,Luscher:2010iy} has gauge-invariant
meaning, and its renormalizability~\cite{Luscher:2011bx} is also quite
suggestive to ERG. Possible connections between ERG and the gradient flow or
diffusion equations have been studied in~Refs.~\cite{Luscher:2013vga,%
Kagimura:2015via,Yamamura:2015kva,Aoki:2016ohw,Pawlowski:2017rhn,%
Makino:2018rys,Abe:2018zdc,Carosso:2018bmz,Carosso:2018rep,Carosso:2019qpb,%
Matsumoto:2020lha}.

One direct connection between ERG and a diffusion equation may be observed as
follows~\cite{Sonoda:2020vut,Sonoda:2019ibh}. The ERG evolution of the Wilson
action~$S_\tau$ is described by the Wilson--Polchinski
equation~\cite{Wilson:1973jj,Polchinski:1983gv}. For the scalar field theory in
$D$-dimensional spacetime, in \emph{dimensionless\/} variables, it
reads\footnote{In momentum space, we adopt the convention
\begin{equation}
   \int_p\equiv\int\frac{d^Dp}{(2\pi)^D},\qquad
   \delta(p)\equiv(2\pi)^D\delta^{(D)}(p),\qquad
   \frac{\delta\phi(p)}{\delta\phi(q)}\equiv\delta(p-q).
\label{eq:(1.1)}
\end{equation}
}
\begin{align}
   \frac{\partial}{\partial\tau}
   e^{S_\tau[\phi]}
   &=\int_p\biggl(
   \left\{
   \left[\frac{\Delta(p)}{K(p)}+\frac{D+2}{2}
   -\frac{\eta_\tau}{2}\right]
   \phi(p)
   +p\cdot\frac{\partial}{\partial p}\phi(p)
   \right\}
   \frac{\delta}{\delta\phi(p)}
\notag\\
   &\qquad\qquad{}
   +\frac{1}{p^2}\left[
   2\frac{\Delta(p)}{K(p)}k(p)
   +2p^2\frac{dk(p)}{dp^2}
   -\eta_\tau k(p)
   \right]
   \frac{1}{2}
   \frac{\delta^2}{\delta\phi(p)\delta\phi(-p)}
   \biggr)\,
   e^{S_\tau[\phi]},
\label{eq:(1.2)}
\end{align}
where $\tau$~parametrizes the ERG evolution and the functions $K(p)$ and~$k(p)$
specify the ERG transformation;
$\Delta(p)\equiv-2p^2(\partial/\partial p^2)K(p)$.

As pointed out in~Ref.~\cite{Sonoda:2015bla}, the ERG evolution of the Wilson
action~$S_\tau$ under~Eq.~\eqref{eq:(1.2)} can be neatly formulated as an
\emph{equality},
\begin{equation}
   \left\langle\!\left\langle
   \phi(p_1)\dotsb\phi(p_n)
   \right\rangle\!\right\rangle_{S_\tau}^{K,k}
   =e^{-\tau n(D+2)/2}Z_\tau^{n/2}
   \left\langle\!\left\langle
   \phi(p_1e^{-\tau})\dotsb\phi(p_ne^{-\tau})
   \right\rangle\!\right\rangle_{S_{\tau=0}}^{K,k}
\label{eq:(1.3)}
\end{equation}
between the modified correlation functions defined by
\begin{align}
   &\left\langle\!\left\langle
   \phi(p_1)\dotsb\phi(p_n)
   \right\rangle\!\right\rangle_S^{K,k}
\notag\\
   &\equiv
   \prod_{i=1}^n\frac{1}{K(p_i)}
   \left\langle
   \exp\left[-\int_p\frac{k(p)}{p^2}
   \frac{1}{2}\frac{\delta^2}{\delta\phi(p)\delta\phi(-p)}\right]
   \phi(p_1)\dotsb\phi(p_n)
   \right\rangle_S,
\label{eq:(1.4)}
\end{align}
where the correlation function on the right-hand side is the conventional one
with respect to the action~$S$. The anomalous dimension~$\eta_\tau$
in~Eq.~\eqref{eq:(1.2)} and the wave function renormalization factor~$Z_\tau$
in~Eq.~\eqref{eq:(1.3)} are related by
\begin{equation}
   \eta_\tau=\frac{\partial}{\partial\tau}\ln Z_\tau.
\label{eq:(1.5)}
\end{equation}
Equation~\eqref{eq:(1.3)} shows that the field variable is multiplicatively
renormalized by~$Z_\tau$ under the ERG evolution, when it is viewed \emph{in
terms of the modified correlation function}. In this sense, what is suitable to
characterize the scaling or critical behavior under the ERG transformation is
the modified correlation function rather than the conventional correlation
function. This fact explains why in~Eq.~\eqref{eq:(1.2)} the anomalous
dimension~$\eta_\tau$, which is related to a ``rescaling'' of the field
variable, should appear not only in the coefficient of the first-order
functional derivative, but also in the coefficient of the second-order
functional derivative. On this issue,
see~Refs.~\cite{Bervillier:2004mf,Igarashi:2016qdr}.

Now, it can be readily seen that Eq.~\eqref{eq:(1.3)} in coordinate
space\footnote{We define
\begin{equation}
   \phi(x)\equiv\int_p e^{ipx}\phi(p),\qquad\phi(p)=\int d^Dx\,e^{-ipx}\phi(x).
\label{eq:(1.6)}
\end{equation}
} is represented in terms of a functional integral as
\begin{align}
   e^{S_\tau[\phi]}
   &=\exp\left[
   \int d^Dx\frac{1}{2}
   \frac{\delta^2}{\delta\phi(x)\delta\phi(x)}\right]
\notag\\
   &\qquad{}
   \times\int[d\phi']\,
   \prod_{x'}\delta
   \left(\phi(x)
   -e^{\tau(D-2)/2}Z_\tau^{1/2}\varphi'(t,x'e^\tau)\right)
\notag\\
   &\qquad\qquad{}
   \times\exp\left[
   -\int d^Dx''\frac{1}{2}
   \frac{\delta^2}{\delta\phi'(x'')\delta\phi'(x'')}\right]
   e^{S_{\tau=0}[\phi']}.
\label{eq:(1.7)}
\end{align}
Here, we assume a particular form of $K$
and~$k$~\cite{Wilson:1973jj},\footnote{This particular choice is not essential.
In fact, one can find the diffusion equation corresponding
to~Eq.~\eqref{eq:(1.2)} with general $K$ and~$k$~\cite{Matsumoto:2020lha}.}
\begin{equation}
   K(p)=e^{-p^2},\qquad k(p)=p^2.
\label{eq:(1.8)}
\end{equation}
The point is that in Eq.~\eqref{eq:(1.7)}, the field $\varphi'(t,x)$ inside the
delta function is given by the solution of the diffusion equation
\begin{equation}
   \partial_t\varphi'(t,x)=\partial^2\varphi'(t,x),\qquad
   \varphi'(t=0,x)=\phi'(x),
\label{eq:(1.9)}
\end{equation}
where the initial configuration for the diffusion is given by the integration
variable~$\phi'$ in the functional integral in~Eq.~\eqref{eq:(1.7)}; the
dimensionless diffusion or flow time~$t$ and the ERG evolution
parameter~$\tau$ are related as
\begin{equation}
   t\equiv e^{2\tau}-1.
\label{eq:(1.10)}
\end{equation}
In this way, one can directly relate the ERG equation in~Eq.~\eqref{eq:(1.2)}
and the diffusion equation in~Eq.~\eqref{eq:(1.9)}. We note that the structure
of~Eq.~\eqref{eq:(1.7)} is very simple: it consists of exponential functions of
the second-order functional derivative and the delta function which imposes the
equality of the argument of the Wilson action and the diffused field. The
diffused field~$\varphi'$ is rescaled in the normalization
by~$e^{\tau(D-2)/2}Z_\tau^{1/2}$, where $(D-2)/2$ is the canonical mass dimension
of the field, and in the spacetime coordinate as~$x\to xe^\tau$.

Considering the continuum limit around a fixed point of the ERG equation, the
above connection relates the correlation function given by the functional
integral with respect to the Wilson action with a finite momentum
cutoff~$\Lambda$ and the correlation function of the diffused field at the
(dimensionful) diffused or flow time $t=1/\Lambda^2$ with respect to the bare
action (with the parameter renormalization, such as the one considered
in~Ref.~\cite{Capponi:2015ucc})~\cite{Sonoda:2019ibh}. This relation provides
an intuitive understanding~\cite{Sonoda:2019ibh} of the fact that the
renormalization of parameters and the wave function of the diffused elementary
scalar field automatically make the equal-point product of diffused fields
finite; the reason is that the functional integral with respect to the Wilson
action possesses an ultraviolet (UV) cutoff~$\Lambda$. This finiteness is
analogous to a remarkable property~\cite{Luscher:2011bx} of the gauge field
diffused by the Yang--Mills gradient flow. These observations motivated a
proposal of GFERG in the pure Yang--Mills theory in~Ref.~\cite{Sonoda:2020vut}.

In the present paper, we generalize the GFERG in~Ref.~\cite{Sonoda:2020vut}
to vector-like gauge theories containing fermion fields. As a natural
generalization, we can maintain the manifest gauge invariance. For the chiral
symmetry, we have two options: one possible formulation
(see~Appendix~\ref{sec:A}) preserves the conventional form of the chiral
symmetry, while the other simpler formulation presented in~Sect.~\ref{sec:2}
realizes the chiral symmetry in a modified form known as the Ginsparg--Wilson
(GW) relation~\cite{Ginsparg:1981bj}. Our derivation of the GW relation in the
present manifestly gauge-invariant ERG formulation is very simple.
In~Sect.~\ref{sec:3}, to have some idea how the GFERG equation works, we
compute a gauge-invariant local Wilson action in quantum electrodynamics (QED)
to the lowest nontrivial order of perturbation theory. Section~\ref{sec:4} is
devoted to our conclusion. In Appendix~\ref{sec:B} we compute the axial
anomaly in~$D=2$ by using our gauge-invariant local Wilson action obtained
in~Sect.~\ref{sec:3}.
 
\section{GFERG for vector-like gauge theories}
\label{sec:2}
Our idea for the construction of a GFERG equation in vector-like gauge
theories would be almost obvious from the elucidation in the previous section.
Imitating the structure of~Eq.~\eqref{eq:(1.7)}, we define the Wilson action by
\begin{align}
   &e^{S_\tau[A,\psi,\Bar{\psi}]}
\notag\\
   &=
   \exp\left[
   \int d^Dx\,
   \frac{1}{2}
   \frac{\delta^2}{\delta A_\mu^a(x)\delta A_\mu^a(x)}\right]
   \exp\left[\int d^Dx'\,
   \frac{\delta}{\delta\psi(x')}
   \frac{\delta}{\delta\Bar{\psi}(x')}
   \right]
\notag\\
   &\qquad{}
   \times
   \int[dA'd\psi'd\Bar{\psi}']\,
   \prod_{x'',\nu,b}\delta
   \left(A_\nu^b(x'')-e^\tau
   g_\tau^{-1}B_\nu^{\prime b}(t,x''e^\tau)\right)
\notag\\
   &\qquad\qquad{}
   \times
   \delta
   \left(
   \psi(x'')-e^{\tau(D-1)/2}Z_\tau^{1/2}\chi'(t,x''e^\tau)
   \right)
   \delta
   \left(
   \Bar{\psi}(x'')-e^{\tau(D-1)/2}Z_\tau^{1/2}\Bar{\chi}'(t,x''e^\tau)
   \right)
\notag\\
   &\qquad\qquad{}
   \times
   \exp\left[-\int d^Dx'''\,
   \frac{\delta}{\delta\psi'(x''')}
   \frac{\delta}{\delta\Bar{\psi}'(x''')}
   \right]
   \exp\left[
   -\int d^Dx''''\,
   \frac{1}{2}
   \frac{\delta^2}
   {\delta A_\rho^{\prime c}(x'''')\delta A_\rho^{\prime c}(x'''')}\right]
\notag\\
   &\qquad\qquad{}
   \times
   \,e^{S_{\tau=0}[A',\psi',\Bar{\psi}']}.
\label{eq:(2.1)}
\end{align}
In this expression, the diffused gauge field~$B'(t,x)$ is the
solution to the Yang--Mills gradient flow equation~\cite{Narayanan:2006rf,
Luscher:2009eq,Luscher:2010iy}
\begin{equation}
   \partial_tB_\mu^{\prime a}(t,x)=D_\nu G_{\nu\mu}^{\prime a}(t,x)
   +\alpha_0 D_\mu\partial_\nu B_\nu^{\prime a}(t,x),\qquad
   B_\mu^{\prime a}(t=0,x)=A_\mu^{\prime a}(x),
\label{eq:(2.2)}
\end{equation}
where $\alpha_0$ is a parameter and the initial configuration~$A'$ is given by
the integration variable in~Eq.~\eqref{eq:(2.1)}. We have defined
\begin{align}
   G_{\mu\nu}^a(t,x)
   &\equiv\partial_\mu B_\nu^a(t,x)-\partial_\nu B_\mu^a(t,x)
   +f^{abc}B_\mu^b(t,x)B_\nu^c(t,x),
\notag\\
   D_\mu X^a(t,x)&\equiv\partial_\mu X^a(t,x)+f^{abc}B_\mu^b(t,x)X^c(t,x)
\label{eq:(2.3)}
\end{align}
from the structure constants of the gauge group $f^{abc}$ defined from
anti-Hermitian generators~$T^a$ by~$[T^a,T^b]=f^{abc}T^c$.
In~Eq.~\eqref{eq:(2.1)} we have taken the canonical mass dimension of the
gauge potential~$1$ and written the wave function renormalization factor of the
gauge field as~$g_\tau^{-2}$; the reason for this convention will become clear
later.\footnote{This convention and the convention
in~Ref.~\cite{Sonoda:2020vut} (especially that in Appendix~A) are related
by~$g_\tau=\lambda z(\tau)$.} Similarly, for the fermion field, we use the
diffusion equations in~Ref.~\cite{Luscher:2013cpa},
\begin{align}
   &\partial_t\chi'(t,x)
   =\left[\Delta-\alpha_0\partial_\mu B_\mu^{\prime a}(t,x)T^a\right]
   \chi'(t,x),&
   \chi'(t=0,x)&=\psi'(x),
\notag\\
   &\partial_t\Bar{\chi}'(t,x)
   =\Bar{\chi}'(t,x)
   \left[\overleftarrow{\Delta}
   +\alpha_0\partial_\mu B_\mu'(t,x)\right],&
   \Bar{\chi}'(t=0,x)&=\Bar{\psi}(x),
\label{eq:(2.4)}
\end{align}
where
\begin{align}
   \Delta&\equiv
   D_\mu D_\mu,&D_\mu&\equiv\partial_\mu+B_\mu^{\prime a}T^a,
\notag\\
   \overleftarrow{\Delta}&\equiv
   \overleftarrow{D}_\mu\overleftarrow{D}_\mu,
   &\overleftarrow{D}_\mu&\equiv\overleftarrow{\partial}_\mu-B_\mu^{\prime a}T^a.
\label{eq:(2.5)}
\end{align}

As in~Ref.~\cite{Sonoda:2020vut}, it is easy to see that the
construction in~Eq.~\eqref{eq:(2.1)} preserves the partition function:
\begin{equation}
   Z=\int[dAd\psi d\Bar{\psi}]\,e^{S_\tau[A,\psi,\Bar{\psi}]}
   =\int[dAd\psi d\Bar{\psi}]\,e^{S_{\tau=0}[A,\psi,\Bar{\psi}]}.
\label{eq:(2.6)}
\end{equation}

Let us examine other properties that follow from~Eq.~\eqref{eq:(2.1)}.

\subsection{Gauge symmetry}
\label{sec:2.1}
In an almost identical way to~Ref.~\cite{Sonoda:2020vut}, we can see that
the Wilson action in~Eq.~\eqref{eq:(2.1)} is invariant under the infinitesimal
gauge transformation
\begin{align}
   A_\mu^a(x)&\to A_\mu^a(x)
   +g_\tau^{-1}\partial_\mu\omega^a(x)
   +f^{abc}A_\mu^b(x)\omega^c(x),
\notag\\
   \psi(x)&\to\psi(x)-\omega^a(x)T^a\psi(x),
\notag\\
   \Bar{\psi}(x)&\to\Bar{\psi}(x)+\Bar{\psi}(x)\omega^a(x)T^a
\label{eq:(2.7)}
\end{align}
if the initial action~$S_{\tau=0}[A,\psi,\Bar{\psi}]$ is invariant under the
above transformation with~$\tau=0$: First, the exponential functions of the
second-order functional derivatives
\begin{equation}
   \exp\left[
   \int d^Dx\,
   \frac{1}{2}
   \frac{\delta^2}{\delta A_\mu^a(x)\delta A_\mu^a(x)}\right]
   \exp\left[\int d^Dx'\,
   \frac{\delta}{\delta\psi(x')}
   \frac{\delta}{\delta\Bar{\psi}(x')}
   \right]
\label{eq:(2.8)}
\end{equation}
are manifestly invariant under the gauge transformation
(see~Ref.~\cite{Sonoda:2020vut}). Then, the gauge transformation on the
argument of the Wilson action in~Eq.~\eqref{eq:(2.1)} is transmitted, through
the delta functions, to the gauge transformation on the diffused fields $B'$,
$\chi'$, and~$\Bar{\chi}'$. This gauge transformation is then, through the
gauge covariance of the diffusion equations, transmitted to that on the initial
configurations $A'$, $\psi'$, and~$\Bar{\psi}'$. Thus, the gauge invariance of
the Wilson action finally depends on the gauge invariance of the initial
action~$S_{\tau=0}[A',\psi',\Bar{\psi}']$ and of the integration
measure~$[dA'd\psi'd\Bar{\psi}']$ (for which we assume its invariance).

In a similar manner, we can see the independence of~$S_\tau[A,\psi,\Bar{\psi}]$
from the parameter~$\alpha_0$ in the diffusion equations,
Eqs.~\eqref{eq:(2.2)} and~\eqref{eq:(2.4)}. To see this, let us suppose that we
make an infinitesimal change of the parameter,
$\alpha_0\to\alpha_0+\delta\alpha_0$. For a fixed initial configuration $A'$,
$\psi'$, and~$\Bar{\psi}'$, the solution of Eqs.~\eqref{eq:(2.2)}
and~\eqref{eq:(2.4)} will change under this. On the other hand, we see that if
we make the following infinitesimal transformation in Eqs.~\eqref{eq:(2.2)}
and~\eqref{eq:(2.4)},
\begin{align}
   B_\mu^a(t,x)&\to B_\mu^a(t,x)+D_\mu\omega^a(t,x),
\notag\\
   \chi(t,x)&\to\chi(t,x)-\omega^a(t,x)T^a\chi(t,x),
\notag\\
   \Bar{\chi}(t,x)&\to\Bar{\chi}(t,x)+\Bar{\chi}(t,x)\omega^a(t,x)T^a,
\label{eq:(2.9)}
\end{align}
where the function~$\omega^a(t,x)$ is defined as the solution of
\begin{equation}
   (\partial_t-\alpha_0D_\nu\partial_\nu)\omega^a(t,x)
   =\delta\alpha_0\partial_\nu B_\nu^a(t,x),
\label{eq:(2.10)}
\end{equation}
then the change of the parameter~$\delta\alpha_0$ in the diffusion equations
can be compensated. By integrating Eq.~\eqref{eq:(2.10)} ``backward against
time'' from~$\omega(t,x)=0$ to~$\omega(t=0,x)$, we then have a gauge
transformation~$\omega(t=0,x)$ on the initial configuration~$A'$, $\psi'$,
and~$\Bar{\psi}'$ such that the solution, $B'$, $\chi'$, and~$\Bar{\chi}'$, is
identical to that before the change of~$\alpha_0$. This shows that if the
initial action~$S_{\tau=0}[A',\psi',\Bar{\psi}']$ and the integration measure
in~Eq.~\eqref{eq:(2.1)} are gauge invariant, then the Wilson
action~$S_\tau[A,\psi,\Bar{\psi}]$ is independent of the parameter~$\alpha_0$.

\subsection{Modified chiral symmetry: GW relation}
\label{sec:2.2}
An important symmetry in a system containing the fermion field is the chiral
symmetry. The Wilson action in~Eq.~\eqref{eq:(2.1)} cannot be invariant under
the conventional form of the chiral transformation, i.e.
\begin{equation}
   \psi(x)\to(1+i\alpha\gamma_5)\psi(x),\qquad
   \Bar{\psi}(x)\to\Bar{\psi}(x)(1+i\alpha\gamma_5).
\label{eq:(2.11)}
\end{equation}
This follows from the fact that, under~Eq.~\eqref{eq:(2.11)},
\begin{equation}
   \frac{\delta}{\delta\psi(x)}
   \to\frac{\delta}{\delta\psi(x)}(1+i\alpha\gamma_5),\qquad
   \frac{\delta}{\delta\Bar{\psi}(x)}
   \to(1+i\alpha\gamma_5)\frac{\delta}{\delta\Bar{\psi}(x)},
\label{eq:(2.12)}
\end{equation}
and thus the exponential function in~Eq.~\eqref{eq:(2.1)},
\begin{equation}
   \Hat{s}\equiv\exp\left[\int d^Dx\,
   \frac{\delta}{\delta\psi(x)}
   \frac{\delta}{\delta\Bar{\psi}(x)}
   \right]
\label{eq:(2.13)}
\end{equation}
does not possess a simple transformation property under~Eq.~\eqref{eq:(2.11)}.
One can avoid this drawback by putting an odd number of Dirac matrices in the
expression such as
\begin{equation}
   \exp\left[-i\int d^Dx\,
   \frac{\delta}{\delta\psi(x)}\Slash{D}_\tau
   \frac{\delta}{\delta\Bar{\psi}(x)}
   \right],
\label{eq:(2.14)}
\end{equation}
where we have to also put the covariant derivative
\begin{equation}
   \Slash{D}_\tau\equiv
   \gamma_\mu\left(\partial_\mu+g_\tau A_\mu^aT^a\right)
\label{eq:(2.15)}
\end{equation}
to preserve the gauge (and Lorentz) invariance. This ``manifestly
chiral-invariant formulation'' is actually a possible option, and we write down
the corresponding ERG equation in~Appendix~\ref{sec:A}.

Here, we pursue the simpler construction, Eq.~\eqref{eq:(2.1)}. Quite
interestingly, the Wilson action in~Eq.~\eqref{eq:(2.1)} can be invariant under
a modified chiral transformation; this is nothing but the chiral symmetry
realized by the GW relation~\cite{Ginsparg:1981bj}\footnote{We would like to
thank Tetsuya Onogi for calling our attention to this point.} (for developments
in the context of lattice gauge theory, see
Refs.~\cite{Neuberger:1997fp,Hasenfratz:1997ft,Neuberger:1998wv,%
Hasenfratz:1998ri,Luscher:1998pqa,Hasenfratz:1998jp}; for studies in the
context of ERG, we may refer, for instance, to
Refs.~\cite{Igarashi:1999rm,Igarashi:2002ba}).

To find the exact chiral symmetry in~Eq.~\eqref{eq:(2.1)}, we introduce
differential operators,
\begin{align}
   \Hat{\gamma}_5&\equiv
   \int d^Dx\,
   \left[
   \gamma_5\psi(x)\frac{\delta}{\delta\psi(x)}
   +\Bar{\psi}(x)\gamma_5\frac{\delta}{\delta\Bar{\psi}(x)}
   \right],
\notag\\
   \Hat{\mit\Gamma}_5&\equiv\Hat{s}\Hat{\gamma}_5\Hat{s}^{-1},
\label{eq:(2.16)}
\end{align}
where $\Hat{s}$ is given in~Eq.~\eqref{eq:(2.13)}. Then,
from~Eq.~\eqref{eq:(2.1)}, we have
\begin{align}
   &\Hat{\mit\Gamma}_5\,e^{S_\tau[A,\psi,\Bar{\psi}]}
\notag\\
   &=
   \exp\left[
   \int d^Dx\,
   \frac{1}{2}
   \frac{\delta^2}{\delta A_\mu^a(x)\delta A_\mu^a(x)}\right]
   \exp\left[\int d^Dx'\,
   \frac{\delta}{\delta\psi(x')}
   \frac{\delta}{\delta\Bar{\psi}(x')}
   \right]
\notag\\
   &\qquad{}
   \times
   \int[dA'd\psi'd\Bar{\psi}']\,
   \prod_{x'',\nu,b}\delta
   \left(A_\nu^b(x'')-e^\tau
   g_\tau^{-1}B_\nu^{\prime b}(t,x''e^\tau)\right)
\notag\\
   &\qquad\qquad{}
   \Hat{\gamma}_5
   \delta
   \left(
   \psi(x'')-e^{\tau(D-1)/2}Z_\tau^{1/2}\chi'(t,x''e^\tau)
   \right)
   \delta
   \left(
   \Bar{\psi}(x'')-e^{\tau(D-1)/2}Z_\tau^{1/2}\Bar{\chi}'(t,x''e^\tau)
   \right)
\notag\\
   &\qquad\qquad{}
   \times
   \exp\left[-\int d^Dx'''\,
   \frac{\delta}{\delta\psi'(x''')}
   \frac{\delta}{\delta\Bar{\psi}'(x''')}
   \right]
   \exp\left[
   -\int d^Dx''''\,
   \frac{1}{2}
   \frac{\delta^2}
   {\delta A_\rho^{\prime c}(x'''')\delta A_\rho^{\prime c}(x'''')}\right]
\notag\\
   &\qquad\qquad{}
   \times
   \,e^{S_{\tau=0}[A',\psi',\Bar{\psi}']}.
\label{eq:(2.17)}
\end{align}
In this expression, $\Hat{\gamma}_5$ acting on $\psi$ and~$\Bar{\psi}$ amounts,
through the delta functions, to the chiral transformation on $\chi'$
and~$\Bar{\chi}'$ because, e.g.,
$\delta(\gamma_5\psi-e^{\tau(D-1)/2}Z_\tau^{1/2}\chi')=
\delta(\psi-e^{\tau(D-1)/2}Z_\tau^{1/2}\gamma_5\chi')$. Since the flow
equations in~Eq.~\eqref{eq:(2.4)} preserve the conventional chiral symmetry,
the chiral transformation on $\chi'$ and~$\Bar{\chi}'$ induces the
transformation on the initial configuration, $\psi'$ and~$\Bar{\psi}'$. Then,
again using the definition of~$\Hat{\mit\Gamma}_5$, we see that\footnote{Here,
we assume that the functional measure $[d\psi'd\Bar{\psi}']$ is invariant under
the conventional chiral transformation.}
\begin{align}
   &\Hat{\mit\Gamma}_5\,e^{S_\tau[A,\psi,\Bar{\psi}]}
\notag\\
   &=
   \exp\left[
   \int d^Dx\,
   \frac{1}{2}
   \frac{\delta^2}{\delta A_\mu^a(x)\delta A_\mu^a(x)}\right]
   \exp\left[\int d^Dx'\,
   \frac{\delta}{\delta\psi(x')}
   \frac{\delta}{\delta\Bar{\psi}(x')}
   \right]
\notag\\
   &\qquad{}
   \times
   \int[dA'd\psi'd\Bar{\psi}']\,
   \prod_{x'',\nu,b}\delta
   \left(A_\nu^b(x'')-e^\tau
   g_\tau^{-1}B_\nu^{\prime b}(t,x''e^\tau)\right)
\notag\\
   &\qquad\qquad{}
   \delta
   \left(
   \psi(x'')-e^{\tau(D-1)/2}Z_\tau^{1/2}\chi'(t,x''e^\tau)
   \right)
   \delta
   \left(
   \Bar{\psi}(x'')-e^{\tau(D-1)/2}Z_\tau^{1/2}\Bar{\chi}'(t,x''e^\tau)
   \right)
\notag\\
   &\qquad\qquad{}
   \times
   \exp\left[-\int d^Dx'''\,
   \frac{\delta}{\delta\psi'(x''')}
   \frac{\delta}{\delta\Bar{\psi}'(x''')}
   \right]
   \exp\left[
   -\int d^Dx''''\,
   \frac{1}{2}
   \frac{\delta^2}
   {\delta A_\rho^{\prime c}(x'''')\delta A_\rho^{\prime c}(x'''')}\right]
\notag\\
   &\qquad\qquad{}
   \times
   \,\Hat{\mit\Gamma}_5\,e^{S_{\tau=0}[A',\psi',\Bar{\psi}']}.
\label{eq:(2.18)}
\end{align}
This shows that
\begin{equation}
   \Hat{\mit\Gamma}_5\,e^{S_{\tau=0}[A,\psi,\Bar{\psi}]}=0\Rightarrow
   \Hat{\mit\Gamma}_5\,e^{S_\tau[A',\psi',\Bar{\psi}']}=0.
\label{eq:(2.19)}
\end{equation}
In this sense, our ERG evolution preserves the invariance under the modified
chiral transformation generated by~$\Hat{\mit\Gamma}_5$.

We note that, from the definition,
\begin{equation}
   \Hat{\mit\Gamma}_5
   =\int d^Dx\,
   \left\{
   \gamma_5\left[
   \psi(x)-\frac{\delta}{\delta\Bar{\psi}(x)}
   \right]\frac{\delta}{\delta\psi(x)}
   +\left[
   \Bar{\psi}(x)+\frac{\delta}{\delta\psi(x)}
   \right]\gamma_5\frac{\delta}{\delta\Bar{\psi}(x)}
   \right\}.
\label{eq:(2.20)}
\end{equation}
From this, we have
\begin{align}
   e^{-S_\tau}\Hat{\mit\Gamma}_5e^{S_\tau}
   &=\int d^Dx\,
   \Biggl\{
   S_\tau\frac{\overleftarrow{\delta}}{\delta\psi(x)}\gamma_5\psi(x)
   +\Bar{\psi}(x)\gamma_5\frac{\delta}{\delta\Bar{\psi}(x)}S_\tau
   -2S_\tau\frac{\overleftarrow{\delta}}{\delta\psi(x)}
   \gamma_5\frac{\delta}{\delta\Bar{\psi}(x)}S_\tau
\notag\\
   &\qquad\qquad\qquad{}
   +2\tr\left[\gamma_5\frac{\delta}{\delta\Bar{\psi}(x)}
   S_\tau\frac{\overleftarrow{\delta}}{\delta\psi(x)}\right]
   \Biggr\}.
\label{eq:(2.21)}
\end{align}
This shows that, if we assume that the action is bilinear in the fermion
field, $S_\tau=-\int d^Dx\,\Bar{\psi}(x)D\psi(x)+\dotsb$, the modified chiral
symmetry of the Wilson action $\Hat{\mit\Gamma}_5\,e^{S_\tau}=0$ implies the GW
relation~\cite{Ginsparg:1981bj},
\begin{equation}
   \gamma_5D+D\gamma_5+2D\gamma_5D=0.
\label{eq:(2.22)}
\end{equation}
Note that we have arrived at this relation by a very simple manipulation while
maintaining a manifest gauge invariance; it would be interesting to see how
this relation reproduces the axial anomaly in our GFERG
formulation.\footnote{In~Appendix~\ref{sec:B} we demonstrate that the
gauge-invariant local Wilson action to~$O(A)$ to be obtained
in~Sect.~\ref{sec:3} actually reproduces the axial anomaly in~$D=2$ correctly.}

\subsection{GFERG equation}
\label{sec:2.3}
Let us derive an ERG equation that the Wilson action in~Eq.~\eqref{eq:(2.1)}
fulfills. This is readily obtained by taking the $\tau$~derivative
of~Eq.~\eqref{eq:(2.1)} in a way analogous to the derivation of the ERG
equation in the Yang--Mills theory~\cite{Sonoda:2020vut}. The result is
\begin{align}
   &\frac{\partial}{\partial\tau}
   e^{S_\tau[A,\psi,\Bar{\psi}]}
\notag\\
   &=\int d^Dx\,
   \frac{\delta}{g_\tau\delta A_\mu^a(x)}
\notag\\
   &\qquad{}
   \times
   \left.\left[
   -2D_\nu F_{\nu\mu}^a(x)-2\alpha_0D_\mu\partial_\nu A_\nu^a(x)
   -\left(\frac{D-2}{2}
   +\frac{\zeta_\tau}{2}
   +x\cdot\frac{\partial}{\partial x}\right)A_\mu^a(x)
   \right]\right|_{A\to g_\tau(A+\delta/\delta A)}
\notag\\
   &\qquad\qquad\qquad{}
   \times
   e^{S_\tau[A,\psi,\Bar{\psi}]}
\notag\\
   &\qquad{}
   +\int d^Dx\,
   \tr\Biggl\{
   \left.\left[
   2\Delta-2\alpha_0\partial_\mu A_\mu(x)
   +\left(\frac{D-1}{2}+\frac{\eta_\tau}{2}+x\cdot\frac{\partial}{\partial x}
   \right)
   \right]\right|_{A\to g_\tau(A+\delta/\delta A)}
\notag\\
   &\qquad\qquad\qquad\qquad{}
   \times\left[\psi(x)-\frac{\delta}{\delta\Bar{\psi}(x)}\right]
   e^{S_\tau[A,\psi,\Bar{\psi}]}\frac{\overleftarrow{\delta}}{\delta\psi(x)}
   \Biggr\}
\notag\\
   &\qquad{}
   +\int d^Dx\,
   \tr\Biggl\{
   \frac{\delta}{\delta\Bar{\psi}(x)}e^{S_\tau[A,\psi,\Bar{\psi}]}
   \left[\Bar{\psi}(x)-\frac{\overleftarrow{\delta}}{\delta\psi(x)}\right]
\notag\\
   &\qquad\qquad\qquad{}
   \times
   \left.\left[
   2\overleftarrow{\Delta}
   +2\alpha_0\partial_\mu A_\mu(x)
   +\left(\frac{D-1}{2}+\frac{\eta_\tau}{2}
   +\frac{\overleftarrow{\partial}}{\partial x}\cdot x\right)
   \right]\right|_{A\to g_\tau(A+\overleftarrow{\delta}/\delta A)}
   \Biggr\}.
\label{eq:(2.23)}
\end{align}
Here, we have defined the anomalous dimensions by (recall~Eq.~\eqref{eq:(1.5)})
\begin{align}
   \zeta_\tau&\equiv4-D-\frac{d}{d\tau}\ln g_\tau^2,
\notag\\
   \eta_\tau&\equiv\frac{d}{d\tau}\ln Z_\tau.
\label{eq:(2.24)}
\end{align}

The ERG equation in~Eq.~\eqref{eq:(2.23)} is the main result of the present
paper. Once this GFERG equation has been obtained, we may forget about the 
underlying construction in~Eq.~\eqref{eq:(2.1)}. Possible requirements on the
initial action~$S_{\tau=0}$ in~Eq.~\eqref{eq:(2.1)} discussed so far, such as
the gauge invariance and the chiral invariance, become implicit. If these
properties of the Wilson action are considered to be desirable, we should
simply pick up a solution or the initial condition of the ERG equation which
fulfills these and other physical requirements (especially the locality and the
Lorentz invariance). In this way, the issue of the existence of the UV
regularization which makes the initial action finite becomes irrelevant. The
renormalizability, i.e.\ whether we can tune parameters in the solution such
that the correlation functions become finite in the continuum limit, is another
issue, and we think that the results
in~Refs.~\cite{Luscher:2011bx,Luscher:2013cpa} become helpful in considering
this question.

Since in~Eq.~\eqref{eq:(2.23)} the power of the gauge potential always
accompanies the power of~$g_\tau$, we see that $g_\tau$ plays the role of the
gauge coupling as the convention indicates. This parameter can thus be used as
an expansion parameter which defines the perturbative expansion at the Gaussian
fixed point~\cite{Sonoda:2020vut}.

\section{Perturbative solution in QED to~$O(g_\tau^1)$}
\label{sec:3}
To have some idea how the GFERG equation in~Eq.~\eqref{eq:(2.23)} works, in
this section we consider the $U(1)$ gauge theory with a Dirac fermion with the
charge~$e$, i.e.
\begin{equation}
  T^a\to-ie,
\label{eq:(3.1)}
\end{equation}
and solve the GFERG equation to the lowest nontrivial order of perturbation
theory, $O(g_\tau^1)$.

We first note that QED possesses charge conjugation symmetry, i.e.\
invariance under\footnote{Although for QED the index~$a$ runs only over~$a=1$,
we keep this index for potential applications of the present lowest-order
solution to non-Abelian theories.}
\begin{equation}
   \psi(x)\to C\bar{\psi}^{\text{T}}(x),
   \qquad\Bar{\psi}(x)\to-\psi^{\text{T}}(x)C^{-1},\qquad
   A_\mu^a(x)\to-A_\mu^a(x),
\label{eq:(3.2)}
\end{equation}
where the charge conjugation matrix satisfies
$C^{-1}\gamma_\mu C=-\gamma_\mu^{\text{T}}$. Since all elements
in~Eq.~\eqref{eq:(2.1)}, especially the flow equations for QED (i.e.\
$f^{abc}=0$), preserve the invariance under~Eq.~\eqref{eq:(3.2)}, if the initial
action~$S_{\tau=0}[A',\psi',\Bar{\psi}']$ is invariant under the charge
conjugation, then $S_\tau[A,\psi,\Bar{\psi}]$ is too. In particular, we can
forbid terms purely consisting of an odd number of gauge potentials; this is
Furry's theorem in the present ERG formulation. Taking this fact into account,
we set the Wilson action as
\begin{align}
   S_\tau[A,\psi,\Bar{\psi}]
   &=\frac{1}{2}\int_pA_\mu^a(-p)T(\tau;p)(p^2\delta_{\mu\nu}-p_\mu p_\nu)
   A_\nu^a(p)
\notag\\
   &\qquad{}
   -\int_p\Bar{\psi}(-p)G(\tau;p)\psi(p)
\notag\\
   &\qquad{}
   +g_\tau
   \int_{p_1,p_2,p_3}\delta(p_1+p_2+p_3)
   \Bar{\psi}(p_1)
   H_\mu^a(\tau;p_1,p_2,p_3)A_\mu^a(p_2)\psi(p_3)
\notag\\
   &\qquad{}
   +O(g_\tau^2).
\label{eq:(3.3)}
\end{align}
For the first term, we already imposed the gauge invariance in~$O(g_\tau^{-1})$,
i.e.\ the invariance under~$A_\mu^a(p)\to
A_\mu^a(p)+g_\tau^{-1}ip_\mu\omega^a(p)$. Note that the function~$G(\tau;p)$ is
not necessarily invariant under~$p\to-p$, because it may contain the Dirac
matrix such as~$\Slash{p}$.

In momentum space, the ERG equation in~Eq.~\eqref{eq:(2.23)} times $e^{-S_\tau}$
reads, when~$\alpha_0=1$,
\begin{align}
   &\frac{\partial}{\partial\tau}S_\tau
\notag\\
   &=\int_p
   \left(2p^2+\frac{D}{2}+1-\frac{\zeta_\tau}{2}
   +p\cdot\frac{\partial}{\partial p}\right)A_\mu^a(p)
   \cdot\frac{\delta S_\tau}{\delta A_\mu^a(p)}
   +\int_p\left(2p^2+1-\frac{\zeta_\tau}{2}\right)
   \frac{\delta S_\tau}{\delta A_\mu^a(p)}
   \frac{\delta S_\tau}{\delta A_\mu^a(-p)}
\notag\\
   &\qquad{}
   +\int_p
   S_\tau\frac{\overleftarrow{\delta}}{\delta\psi(p)}
   \left(2p^2+\frac{D}{2}+\frac{1}{2}-\frac{\eta_\tau}{2}
   +p\cdot\frac{\partial}{\partial p}\right)
   \psi(p)
\notag\\
   &\qquad{}
   +\int_p
   \Bar{\psi}(p)\left(2p^2+\frac{D}{2}+\frac{1}{2}-\frac{\eta_\tau}{2}
   +\frac{\overleftarrow{\partial}}{\partial p}\cdot p\right)
   \frac{\delta}{\delta\Bar{\psi}(p)}S_\tau
\notag\\
   &\qquad{}
   +\int_p
   (-1)\left(4p^2+1-\eta_\tau\right)
   S_\tau\frac{\overleftarrow{\delta}}{\delta\psi(p)}   
   \frac{\delta}{\delta\Bar{\psi}(-p)}S_\tau
\notag\\
   &\qquad{}
   +g_\tau
   \int_{p,p',p''}\delta(p+p'+p'')
\notag\\
   &\qquad\qquad{}
   \times
   \Biggr\{
   -4i
   S_\tau\frac{\overleftarrow{\delta}}{\delta\psi(-p)}
   \left[A_\mu^a(p')+\frac{\delta S_\tau}{\delta A_\mu^a(-p')}\right]T^a
   p''_\mu
   \psi(p'')
   \notag\\
   &\qquad\qquad\qquad{}
   +4i
   \Bar{\psi}(p)
   p_\mu
   \left[A_\mu^a(p')+\frac{\delta S_\tau}{\delta A_\mu^a(-p')}\right]T^a
   \frac{\delta}{\delta\Bar{\psi}(-p'')}S_\tau
   \Biggr\}
\notag\\
   &\qquad{}
   +g_\tau
   \int_{p,p',p''}\delta(p+p'+p'')
\notag\\
   &\qquad\qquad{}
   \times
   \tr\Biggr\{
   -4i
   \frac{\delta}{\delta\Bar{\psi}(-p)}S_\tau\cdot
   S_\tau\frac{\overleftarrow{\delta}}{\delta\psi(-p')}
   (p-p')_\mu
   \left[A_\mu^a(p'')+\frac{\delta S_\tau}{\delta A_\mu^a(-p'')}\right]T^a
   \Biggr\}
\notag\\
   &\qquad{}
   +O(g_\tau^2),
\label{eq:(3.4)}
\end{align}
where we have retained only terms relevant to the ERG evolution of terms
in~Eq.~\eqref{eq:(3.3)}.

\subsection{$O(g_\tau^0)$ terms}
\label{sec:3.1}
In the lowest order, $O(g_\tau^0)$, the ERG equation in~Eq.~\eqref{eq:(3.4)}
requires, for the coefficient functions in~Eq.~\eqref{eq:(3.3)},
\begin{align}
   \frac{1}{2}\frac{\partial}{\partial\tau}T(\tau;p)
   &=\left(-\frac{1}{2}p\cdot\frac{\partial}{\partial p}+2p^2-\frac{\zeta_\tau}{2}\right)
   T(\tau;p)
   +p^2\left(2p^2+1-\frac{\zeta_\tau}{2}\right)T(\tau;p)^2,
\notag\\
   \frac{\partial}{\partial\tau}G(\tau;p)
   &=\left(-p\cdot\frac{\partial}{\partial p}
   +4p^2+1-\eta_\tau\right)G(\tau;p)
   +(4p^2+1-\eta_\tau)G(\tau;p)^2.
\label{eq:(3.5)}
\end{align}
It can be seen that the general solutions to these are given by\footnote{For
this, we note that the differential equations in~Eq.~\eqref{eq:(3.5)} become
linear in terms of~$T^{-1}$ and~$G^{-1}$.}
\begin{equation}
   T(\tau;p)=-\frac{1}{e^{\tau(4-D)}g_\tau^{-2}C(e^{-\tau}p)e^{-2p^2}+p^2},\qquad
   G(\tau;p)=
   -\frac{\Slash{p}}{Z_\tau\widetilde{C}(e^{-\tau}p)e^{-2p^2}+\Slash{p}},
\label{eq:(3.6)}
\end{equation}
where $C(p)$ and~$\widetilde{C}(p)$ are arbitrary functions of~$p^2$; for
locality of the Wilson action, however, $C(p)$ and~$\widetilde{C}(p)$ must be
analytic at~$p=0$. In obtaining the above expression for~$G$, we have assumed
parity symmetry and that $\widetilde{C}$ does not contain~$\gamma_5$, and thus
$\Slash{p}$ and~$\widetilde{C}$ commute with each other.

\subsection{GW relation in~$O(g_\tau^0)$}
\label{sec:3.2}
Even in the the above lowest $O(g_\tau^0)$ solution, it is interesting to see
how the GW relation in~Eq.~\eqref{eq:(2.22)} is realized. To this order,
Eq.~\eqref{eq:(2.22)} implies
\begin{equation}
   \gamma_5G+G\gamma_5+2G\gamma_5G=0
   \Leftrightarrow
   G^{-1}\gamma_5+\gamma_5G^{-1}+2\gamma_5=0.
\label{eq:(3.7)}
\end{equation}
For Eq.~\eqref{eq:(3.6)}, on the other hand, we have
\begin{equation}
   G^{-1}\gamma_5+\gamma_5G^{-1}+2\gamma_5
   =\frac{Z_\tau e^{-2p^2}}{\Slash{p}}[\gamma_5,\widetilde{C}(e^{-\tau}p)].
\label{eq:(3.8)}
\end{equation}
Therefore, if and only if $\gamma_5$ and the function~$\widetilde{C}$
in~Eq.~\eqref{eq:(3.6)} commute, i.e.\ if and only if $\widetilde{C}$ does not
contain~$\Slash{p}$, the Wilson action satisfies the GW relation. Note that,
since the $\tau$~dependence of~$G$ arises only from the combination~$e^{-\tau}p$
in~$\widetilde{C}$, the GW relation is preserved under the evolution
of~$\tau$, as our general discussion shows.

An interesting case in which the GW relation is \emph{not\/} fulfilled is
\begin{equation}
   \widetilde{C}(e^{-\tau}p)
   =\frac{e^{-\tau}\Slash{p}}{e^{-\tau}\Slash{p}+im}
   =\frac{\Slash{p}}{\Slash{p}+ie^\tau m},
\label{eq:(3.9)}
\end{equation}
where $m$ is a constant. In this case, the breaking of the GW
relation in~Eq.~\eqref{eq:(3.8)} becomes
\begin{equation}
   G^{-1}\gamma_5+\gamma_5G^{-1}+2\gamma_5
   =2iZ_\tau e^{-2p^2}
   \gamma_5\frac{e^\tau m}{p^2+e^{2\tau}m^2}.
\label{eq:(3.10)}
\end{equation}
The choice of~$\widetilde{C}$ in~Eq.~\eqref{eq:(3.9)} actually realizes a
massive fermion. The propagator of the fermion field with respect to the Wilson
action to this order is given by
\begin{align}
   \left\langle\psi(p)\Bar{\psi}(q)\right\rangle_{S_\tau}
   &=G(\tau;p)^{-1}\delta(p+q)
\notag\\
   &=-\frac{Z_\tau e^{-2p^2}}{\Slash{p}+ie^\tau m}\delta(p+q)-\delta(p+q).
\label{eq:(3.11)}
\end{align}
This is not, however, the propagator that obeys the scaling law under the ERG
evolution; recall the discussion at~Eq.~\eqref{eq:(1.3)}. Such a propagator
is given by the modified one~\cite{Sonoda:2015bla} defined by
(see~Eq.~\eqref{eq:(1.4)} for the scalar field case)
\begin{align}
   \left\langle\!\left\langle\psi(p)\Bar{\psi}(q)
   \right\rangle\!\right\rangle_{S_\tau}
   &\equiv
   e^{p^2}e^{q^2}
   \left\langle
   \exp\left[-\int_r\frac{\delta}{\delta\psi(r)}
   \frac{\delta}{\delta\Bar{\psi}(-r)}\right]
   \psi(p)\Bar{\psi}(q)\right\rangle_{S_\tau}
\notag\\
   &=
   e^{p^2}e^{q^2}
   \left[\left\langle\psi(p)\Bar{\psi}(q)\right\rangle_{S_\tau}
   +\delta(p+q)\right]
\notag\\
   &=-\frac{Z_\tau}{\Slash{p}+ie^\tau m}\delta(p+q).
\label{eq:(3.12)}
\end{align}
The correlation length in units of the UV cutoff is thus given
by~$e^{-\tau}m^{-1}$, and $m$ is the mass parameter; the critical surface is
approached by~$m\to0$. As expected, the GW relation is broken by the amount of
this mass parameter as~Eq.~\eqref{eq:(3.10)}.

The GW relation in~Eq.~\eqref{eq:(3.7)} is satisfied if $\widetilde{C}(p)$ is a
scalar function of~$p^2$ so that $\widetilde{C}$ commutes with~$\gamma_5$. In
this case, the function~$G$ realizes a massless fermion.

\subsection{$O(g_\tau^1)$ terms}
\label{sec:(3.3)}
Next, we consider the $O(g_\tau^1)$ terms. Equation~\eqref{eq:(3.4)} requires,
for the coefficient functions in~Eq.~\eqref{eq:(3.4)},
\begin{align}
   &\left[\frac{\partial}{\partial\tau}
   +\sum_ip_i\cdot\frac{\partial}{\partial p_i}
   -2\sum_ip_i^2
   +\eta_\tau
   \right]H_\mu^a(\tau;p_1,p_2,p_3)
\notag\\
   &\qquad{}
   -(4p_1^2+1-\eta_\tau)G(\tau;-p_1)H_\mu^a(\tau;p_1,p_2,p_3)
   -(4p_3^2+1-\eta_\tau)H_\mu^a(\tau;p_1,p_2,p_3)G(\tau;p_3)
\notag\\
   &\qquad{}
   -2\left(2p_2^2+1-\frac{\zeta_\tau}{2}\right)p_2^2T(\tau;p_2)
   \left(\delta_{\mu\nu}-\frac{p_{2\mu}p_{2\nu}}{p_2^2}\right)
   H_\nu^a(\tau;p_1,p_2,p_3)
\notag\\
   &=4iT^aG(\tau;-p_1)p_{3\mu}
   -4iT^ap_{1\mu}G(\tau;p_3)
   -4iT^aG(\tau;-p_1)(p_1-p_3)_\mu G(\tau;p_3)
\notag\\
   &\qquad{}
   +4iT^ap_2^2T(\tau;p_2)G(\tau;-p_1)
   \left(\delta_{\mu\nu}-\frac{p_{2\mu}p_{2\nu}}{p_2^2}\right)p_{3\nu}
\notag\\
   &\qquad{}
   -4iT^ap_2^2T(\tau;p_2)G(\tau;p_3)
   \left(\delta_{\mu\nu}-\frac{p_{2\mu}p_{2\nu}}{p_2^2}\right)p_{1\nu}
\notag\\
   &\qquad{}
   -4iT^ap_2^2T(\tau;p_2)
   G(\tau;-p_1)
   G(\tau;p_3)
   \left(\delta_{\mu\nu}-\frac{p_{2\mu}p_{2\nu}}{p_2^2}\right)
   (p_1-p_3)_\nu.
\label{eq:(3.13)}
\end{align}
To solve this differential equation we define the decomposition into
transverse and longitudinal parts as\footnote{The following strategy to
solve the ERG equation was obtained through discussions with Hidenori Sonoda,
and we would like to thank him.}
\begin{align}
   H_\mu^a(\tau;p_1,p_2,p_3)
   &=\left(\delta_{\mu\nu}-\frac{p_{2\mu}p_{2\nu}}{p_2^2}\right)
   H_\nu^a(\tau;p_1,p_2,p_3)
   +\frac{p_{2\mu}p_{2\nu}}{p_2^2}
   H_\nu^a(\tau;p_1,p_2,p_3)
\notag\\
   &\equiv
   t_\mu^a(\tau;p_1,p_2,p_3)+\ell_\mu^a(\tau;p_1,p_2,p_3).
\label{eq:(3.14)}
\end{align}
Then, the ERG equation is decomposed into
\begin{align}
   &\left[\frac{\partial}{\partial\tau}
   +\sum_ip_i\cdot\frac{\partial}{\partial p_i}
   -2\sum_ip_i^2
   +\eta_\tau
   \right]\ell_\mu^a(\tau;p_1,p_2,p_3)
\notag\\
   &\qquad{}
   -(4p_1^2+1-\eta_\tau)G(\tau;-p_1)\ell_\mu^a(\tau;p_1,p_2,p_3)
   -(4p_3^2+1-\eta_\tau)\ell_\mu^a(\tau;p_1,p_2,p_3)G(\tau;p_3)
\notag\\
   &=4iT^aG(\tau;-p_1)\frac{p_{2\mu}p_{2\nu}}{p_2^2}p_{3\nu}
   -4iT^aG(\tau;p_3)\frac{p_{2\mu}p_{2\nu}}{p_2^2}p_{1\nu}
\notag\\
   &\qquad{}
   -4iT^aG(\tau;-p_1)G(\tau;p_3)\frac{p_{2\mu}p_{2\nu}}{p_2^2}(p_1-p_3)_\nu,
\label{eq:(3.15)}
\end{align}
and
\begin{align}
   &\left[\frac{\partial}{\partial\tau}
   +\sum_ip_i\cdot\frac{\partial}{\partial p_i}
   -2\sum_ip_i^2
   +\eta_\tau
   \right]t_\mu^a(\tau;p_1,p_2,p_3)
\notag\\
   &\qquad{}
   -(4p_1^2+1-\eta_\tau)G(\tau;-p_1)t_\mu^a(\tau;p_1,p_2,p_3)
   -(4p_3^2+1-\eta_\tau)t_\mu^a(\tau;p_1,p_2,p_3)G(\tau;p_3)
\notag\\
   &\qquad{}
   -2\left(2p_2^2+1-\frac{\zeta_\tau}{2}\right)p_2^2T(\tau;p_2)
   t_\mu^a(\tau;p_1,p_2,p_3)
\notag\\
   &=4iT^a[1+p_2^2T(\tau;p_2)]G(\tau;-p_1)
   \left(\delta_{\mu\nu}-\frac{p_{2\mu}p_{2\nu}}{p_2^2}\right)p_{3\nu}
\notag\\
   &\qquad{}
   -4iT^a[1+p_2^2T(\tau;p_2)]G(\tau;p_3)
   \left(\delta_{\mu\nu}-\frac{p_{2\mu}p_{2\nu}}{p_2^2}\right)p_{1\nu}
\notag\\
   &\qquad{}
   -4iT^a[1+p_2^2T(\tau;p_2)]G(\tau;-p_1)G(\tau;p_3)
   \left(\delta_{\mu\nu}-\frac{p_{2\mu}p_{2\nu}}{p_2^2}\right)(p_1-p_3)_\nu.
\label{eq:(3.16)}
\end{align}

\subsubsection{Solution for~$\ell_\mu$}
First, to solve Eq.~\eqref{eq:(3.15)}, we set
\begin{equation}
   \ell_\mu^a(\tau;p_1,p_2,p_3)
   =e^{-2\tau}Z_\tau e^{-p_1^2}G(\tau;-p_1)
   e^{p_2^2}\Tilde{\ell}_\mu^a(\tau;p_1,p_2,p_3)
   e^{-p_3^2}G(\tau;p_3).
\label{eq:(3.17)}
\end{equation}
Then, noting the relation
\begin{equation}
   \left[
   \frac{\partial}{\partial\tau}
   +p\cdot\frac{\partial}{\partial p}
   -2p^2+\frac{\eta_\tau}{2}
   -(4p^2+1-\eta_\tau)G(\tau;p)
   \right]
   e^{-\tau}Z_\tau^{1/2}e^{-p^2}G(\tau;p)=0,
\label{eq:(3.18)}
\end{equation}
which follows from~Eq.~\eqref{eq:(3.5)}, Eq.~\eqref{eq:(3.15)} reduces to
\begin{align}
   &\left(\frac{\partial}{\partial\tau}
   +\sum_ip_i\cdot\frac{\partial}{\partial p_i}
   \right)\Tilde{\ell}_\mu^a(\tau;p_1,p_2,p_3)
\notag\\
   &=-4iT^ae^{2\tau}e^{-p_1^2-p_2^2+p_3^2}
   \frac{\widetilde{C}(e^{-\tau}p_1)}{\Slash{p}_1}
   \frac{p_{2\mu}p_{2\nu}}{p_2^2}p_{1\nu}
   -4iT^ae^{2\tau}e^{p_1^2-p_2^2-p_3^2}
   \frac{\widetilde{C}(e^{-\tau}p_3)}{\Slash{p}_3}
   \frac{p_{2\mu}p_{2\nu}}{p_2^2}p_{3\nu},
\label{eq:(3.19)}
\end{align}
where we have used Eq.~\eqref{eq:(3.6)}. The general solution to this is given
by
\begin{align}
   \Tilde{\ell}_\mu^a(\tau;p_1,p_2,p_3)
   &=f_\mu^a(e^{-\tau}p_1,e^{-\tau}p_2,e^{-\tau}p_3)
\notag\\
   &\qquad{}
   -4iT^ae^{2\tau}W(p_3;p_2,p_1)
   \frac{\widetilde{C}(e^{-\tau}p_1)}{\Slash{p}_1}
   \frac{p_{2\mu}p_{2\nu}}{p_2^2}p_{1\nu}
\notag\\
   &\qquad{}
   -4iT^ae^{2\tau}W(p_1;p_2,p_3)
   \frac{\widetilde{C}(e^{-\tau}p_3)}{\Slash{p}_3}
   \frac{p_{2\mu}p_{2\nu}}{p_2^2}p_{3\nu},
\label{eq:(3.20)}
\end{align}
where $f_\mu^a(p_1,p_2,p_3)$ is an arbitrary vector function of~$p_i$; in this
expression, the function~$W$ is defined by
\begin{equation}
   W(p_1;p_2,p_3)\equiv\frac{1}{2}
   \frac{e^{p_1^2-p_2^2-p_3^2}-1}{p_1^2-p_2^2-p_3^2},
\label{eq:(3.21)}
\end{equation}
which solves
\begin{equation}
   \left(\frac{\partial}{\partial\tau}
   +\sum_ip_i\cdot\frac{\partial}{\partial p_i}
   \right)e^{2\tau}W(p_1;p_2,p_3)
   =e^{2\tau}e^{p_1^2-p_2^2-p_3^2}.
\label{eq:(3.22)}
\end{equation}

Going back to Eq.~\eqref{eq:(3.17)}, we have the general solution
for~$\ell_\mu^a$,
\begin{align}
   &\ell_\mu^a(\tau;p_1,p_2,p_3)
\notag\\
   &=e^{-2\tau}Z_\tau e^{-p_1^2+p_2^2-p_3^2}G(\tau;-p_1)
   f_\mu^a(e^{-\tau}p_1,e^{-\tau}p_2,e^{-\tau}p_3)
   G(\tau;p_3)
\notag\\
   &\qquad{}
   -4iT^aZ_\tau e^{-p_1^2+p_2^2-p_3^2}W(p_3;p_2,p_1)
   G(\tau;-p_1)\frac{\widetilde{C}(e^{-\tau}p_1)}{\Slash{p}_1}G(\tau;p_3)
   \frac{p_{2\mu}p_{2\nu}}{p_2^2}p_{1\nu}
\notag\\
   &\qquad{}
   -4iT^a
   Z_\tau e^{-p_1^2+p_2^2-p_3^2}W(p_1;p_2,p_3)
   G(\tau;-p_1)\frac{\widetilde{C}(e^{-\tau}p_3)}{\Slash{p}_3}G(\tau;p_3)
   \frac{p_{2\mu}p_{2\nu}}{p_2^2}p_{3\nu}.
\label{eq:(3.23)}
\end{align}

\subsubsection{Solution for~$t_\mu$}
Next, to solve~Eq.~\eqref{eq:(3.16)}, we set
\begin{equation}
   t_\mu^a(\tau;p_1,p_2,p_3)
   =e^{-2\tau}Z_\tau e^{\tau(4-D)}g_\tau^{-2}
   e^{-p_1^2}G(\tau;-p_1)e^{-p_2^2}T(\tau;p_2)
   \Tilde{t}_\mu^a(\tau;p_1,p_2,p_3)e^{-p_3^2}G(\tau;p_3),
\label{eq:(3.24)}
\end{equation}
in view of
\begin{equation}
   \left[
   \frac{\partial}{\partial\tau}
   +p\cdot\frac{\partial}{\partial p}
   -2p^2
   -2\left(2p^2+1-\frac{\zeta_\tau}{2}\right)p^2T(\tau;p)
   \right]
   e^{\tau(4-D)}g_\tau^{-2}e^{-p^2}T(\tau;p)=0.
\label{eq:(3.25)}
\end{equation}
Then Eq.~\eqref{eq:(3.16)} reduces to
\begin{align}
   &\left(\frac{\partial}{\partial\tau}
   +\sum_ip_i\cdot\frac{\partial}{\partial p_i}
   \right)
   \Tilde{t}_\mu^a(\tau;p_1,p_2,p_3)
\notag\\
   &=4iT^a
   e^{2\tau}e^{-p_1^2-p_2^2+p_3^2}
   \frac{\widetilde{C}(e^{-\tau}p_1)}{\Slash{p}_1}
   C(e^{-\tau}p_2)
   \left(\delta_{\mu\nu}-\frac{p_{2\mu}p_{2\nu}}{p_2^2}\right)p_{1\nu}
\notag\\
   &\qquad{}
   +4iT^a
   e^{2\tau}e^{p_1^2-p_2^2-p_3^2}
   \frac{\widetilde{C}(e^{-\tau}p_3)}{\Slash{p}_3}
   C(e^{-\tau}p_2)
   \left(\delta_{\mu\nu}-\frac{p_{2\mu}p_{2\nu}}{p_2^2}\right)p_{3\nu}.
\label{eq:(3.26)}
\end{align}
The general solution to this is given by
\begin{align}
   \Tilde{t}_\mu^a(\tau;p_1,p_2,p_3)
   &=g_\mu^a(e^{-\tau}p_1,e^{-\tau}p_2,e^{-\tau}p_3)
\notag\\
   &\qquad{}
   +4iT^ae^{2\tau}W(p_3;p_2,p_1)
   \frac{\widetilde{C}(e^{-\tau}p_1)}{\Slash{p}_1}
   C(e^{-\tau}p_2)
   \left(\delta_{\mu\nu}-\frac{p_{2\mu}p_{2\nu}}{p_2^2}\right)p_{1\nu}
\notag\\
   &\qquad{}
   +4iT^ae^{2\tau}W(p_1;p_2,p_3)
   \frac{\widetilde{C}(e^{-\tau}p_3)}{\Slash{p}_3}
   C(e^{-\tau}p_2)
   \left(\delta_{\mu\nu}-\frac{p_{2\mu}p_{2\nu}}{p_2^2}\right)p_{3\nu},
\label{eq:(3.27)}
\end{align}
where $g_\mu^a(p_1,p_2,p_3)$ is an arbitrary vector function of~$p_i$;
$g_\mu^a(p_1,p_2,p_3)$ must, however, be transverse,
$p_{2\mu}g_\mu^a(p_1,p_2,p_3)=0$. Plugging this into~Eq.~\eqref{eq:(3.24)},
we have
\begin{align}
   &t_\mu^a(\tau;p_1,p_2,p_3)
\notag\\
   &=e^{-2\tau}Z_\tau e^{\tau(4-D)}g_\tau^{-2}
   e^{-p_1^2-p_2^2-p_3^2}G(\tau;-p_1)T(\tau;p_2)
   g_\mu^a(e^{-\tau}p_1,e^{-\tau}p_2,e^{-\tau}p_3)G(\tau;p_3)
\notag\\
   &\qquad{}
   +4iT^aZ_\tau e^{\tau(4-D)}g_\tau^{-2}e^{-p_1^2-p_2^2-p_3^2}W(p_3;p_2,p_1)
   G(\tau;-p_1)\frac{\widetilde{C}(e^{-\tau}p_1)}{\Slash{p}_1}
\notag\\
   &\qquad\qquad{}
   \times
   T(\tau;p_2)C(e^{-\tau}p_2)
   G(\tau;p_3)
   \left(\delta_{\mu\nu}-\frac{p_{2\mu}p_{2\nu}}{p_2^2}\right)p_{1\nu}
\notag\\
   &\qquad{}
   +4iT^aZ_\tau e^{\tau(4-D)}g_\tau^{-2}e^{-p_1^2-p_2^2-p_3^2}W(p_1;p_2,p_3)
   G(\tau;-p_1)
   \frac{\widetilde{C}(e^{-\tau}p_3)}{\Slash{p}_3}G(\tau;p_3)
\notag\\
   &\qquad\qquad{}
   \times
   T(\tau;p_2)C(e^{-\tau}p_2)
   \left(\delta_{\mu\nu}-\frac{p_{2\mu}p_{2\nu}}{p_2^2}\right)p_{3\nu}.
\label{eq:(3.28)}
\end{align}

\subsubsection{Gauge invariance}
So far, we have obtained a most general form of the interaction
vertex~$H_\mu^a$ in~Eq.~\eqref{eq:(3.3)}, which solves the ERG equation. We now
impose the gauge invariance to the Wilson action and further
restrict~$H_\mu^a$. The gauge transformation in~Eq.~\eqref{eq:(2.7)} reads, in
momentum space,
\begin{align}
   A_\mu^a(p)&\to A_\mu^a(p)
   +g_\tau^{-1}ip_\mu\omega^a(p),
\notag\\
   \psi(p)&\to
   \psi(p)-\int_q\omega^a(p-q)T^a\psi(q),
\notag\\
   \Bar{\psi}(p)&\to
   \Bar{\psi}(p)+\int_q\Bar{\psi}(q)\omega^a(p-q)T^a.
\label{eq:(3.29)}
\end{align}
The gauge invariance of the Wilson action in~Eq.~\eqref{eq:(3.3)}
to~$O(g_\tau^0)$ thus requires
\begin{equation}
   ip_{2\mu}H_\mu^a(\tau;p_1,p_2,p_3)
   =T^aG(\tau;p_3)-T^aG(\tau;-p_1).
\label{eq:(3.30)}
\end{equation}
From Eqs.~\eqref{eq:(3.14)} and~\eqref{eq:(3.23)}, we thus have (note that the
transverse part in~Eq.~\eqref{eq:(3.28)} does not contribute to this)
\begin{align}
   &ip_{2\mu}H_\mu^a(\tau;p_1,p_2,p_3)
\notag\\
   &=e^{-2\tau}Z_\tau e^{-p_1^2+p_2^2-p_3^2}G(\tau;-p_1)
   ip_{2\mu}f_\mu^a(e^{-\tau}p_1,e^{-\tau}p_2,e^{-\tau}p_3)
   G(\tau;p_3)
\notag\\
   &\qquad{}
   +T^aZ_\tau
   \left(e^{-2p_1^2}-e^{-p_1^2+p_2^2-p_3^2}\right)
   G(\tau;-p_1)\frac{\widetilde{C}(e^{-\tau}p_1)}{\Slash{p}_1}G(\tau;p_3)
\notag\\
   &\qquad{}
   +T^aZ_\tau
   \left(e^{-2p_3^2}-e^{-p_1^2+p_2^2-p_3^2}\right)
   G(\tau;-p_1)\frac{\widetilde{C}(e^{-\tau}p_3)}{\Slash{p}_3}G(\tau;p_3)
\notag\\
   &=e^{-2\tau}Z_\tau e^{-p_1^2+p_2^2-p_3^2}G(\tau;-p_1)
   ip_{2\mu}f_\mu^a(e^{-\tau}p_1,e^{-\tau}p_2,e^{-\tau}p_3)
   G(\tau;p_3)
\notag\\
   &\qquad{}
   -T^aZ_\tau
   e^{-p_1^2+p_2^2-p_3^2}
   G(\tau;-p_1)\frac{\widetilde{C}(e^{-\tau}p_1)}{\Slash{p}_1}G(\tau;p_3)
\notag\\
   &\qquad{}
   -T^aZ_\tau
   e^{-p_1^2+p_2^2-p_3^2}
   G(\tau;-p_1)\frac{\widetilde{C}(e^{-\tau}p_3)}{\Slash{p}_3}G(\tau;p_3)
\notag\\
   &\qquad{}
   +T^aG(\tau;p_3)
   -T^aG(\tau;-p_1),
\label{eq:(3.31)}
\end{align}
where we have used $p_3^2-p_2^2-p_1^2=2p_1\cdot p_2$
and~$p_1^2-p_2^2-p_3^2=2p_2\cdot p_3$, which follow from the momentum
conservation~$p_1+p_2+p_3=0$ and the relation
\begin{equation}
   Z_\tau e^{-2p^2}G(\tau;p)\frac{\widetilde{C}(e^{-\tau}p)}{\Slash{p}}
   =-1-G(\tau;p),
\label{eq:(3.32)}
\end{equation}
which follows from~Eq.~\eqref{eq:(3.6)}. Equations~\eqref{eq:(3.30)}
and~\eqref{eq:(3.31)} show that we can achieve the gauge invariance by
taking
\begin{align}
   f_\mu^a(e^{-\tau}p_1,e^{-\tau}p_2,e^{-\tau}p_3)
   =-iT^a\left[
   \frac{\widetilde{C}(e^{-\tau}p_1)}{e^{-\tau}\Slash{p}_1}
   +\frac{\widetilde{C}(e^{-\tau}p_3)}{e^{-\tau}\Slash{p}_3}
   \right]\frac{e^{-\tau}p_{2\mu}}{e^{-2\tau}p_2^2}.
\label{eq:(3.33)}
\end{align}
Note that the gauge invariance is preserved under the evolution of~$\tau$,
reflecting the gauge invariance of the present ERG formulation.

Therefore, adding Eqs.~\eqref{eq:(3.23)} and~\eqref{eq:(3.28)}, we have
\begin{align}
   &H_\mu^a(\tau;p_1,p_2,p_3)
\notag\\
   &=e^{-2\tau}Z_\tau e^{\tau(4-D)}g_\tau^{-2}
   e^{-p_1^2-p_2^2-p_3^2}G(\tau;-p_1)T(\tau;p_2)
   g_\mu^a(e^{-\tau}p_1,e^{-\tau}p_2,e^{-\tau}p_3)G(\tau;p_3)
\notag\\
   &\qquad{}
   -iT^aZ_\tau e^{-p_1^2+p_2^2-p_3^2}G(\tau;-p_1)
   \left[
   \frac{\widetilde{C}(e^{-\tau}p_1)}{\Slash{p}_1}
   +\frac{\widetilde{C}(e^{-\tau}p_3)}{\Slash{p}_3}
   \right]\frac{p_{2\mu}}{p_2^2}
   G(\tau;p_3)
\notag\\
   &\qquad{}
   -4iT^aZ_\tau e^{-p_1^2+p_2^2-p_3^2}W(p_3;p_2,p_1)
   G(\tau;-p_1)\frac{\widetilde{C}(e^{-\tau}p_1)}{\Slash{p}_1}G(\tau;p_3)
   \frac{p_{2\mu}p_{2\nu}}{p_2^2}p_{1\nu}
\notag\\
   &\qquad{}
   -4iT^a
   Z_\tau e^{-p_1^2+p_2^2-p_3^2}W(p_1;p_2,p_3)
   G(\tau;-p_1)\frac{\widetilde{C}(e^{-\tau}p_3)}{\Slash{p}_3}G(\tau;p_3)
   \frac{p_{2\mu}p_{2\nu}}{p_2^2}p_{3\nu}.
\notag\\
   &\qquad{}
   +4iT^aZ_\tau e^{\tau(4-D)}g_\tau^{-2}e^{-p_1^2-p_2^2-p_3^2}W(p_3;p_2,p_1)
   G(\tau;-p_1)\frac{\widetilde{C}(e^{-\tau}p_1)}{\Slash{p}_1}
\notag\\
   &\qquad\qquad{}
   \times
   T(\tau;p_2)C(e^{-\tau}p_2)
   G(\tau;p_3)
   \left(\delta_{\mu\nu}-\frac{p_{2\mu}p_{2\nu}}{p_2^2}\right)p_{1\nu}
\notag\\
   &\qquad{}
   +4iT^aZ_\tau e^{\tau(4-D)}g_\tau^{-2}
   e^{-p_1^2-p_2^2-p_3^2}W(p_1;p_2,p_3)
   G(\tau;-p_1)
\notag\\
   &\qquad\qquad{}
   \times
   T(\tau;p_2)C(e^{-\tau}p_2)
   \frac{\widetilde{C}(e^{-\tau}p_3)}{\Slash{p}_3}G(\tau;p_3)
   \left(\delta_{\mu\nu}-\frac{p_{2\mu}p_{2\nu}}{p_2^2}\right)p_{3\nu}
\notag\\
   &=e^{-2\tau}Z_\tau e^{\tau(4-D)}g_\tau^{-2}
   e^{-p_1^2-p_2^2-p_3^2}G(\tau;-p_1)T(\tau;p_2)
   g_\mu^a(e^{-\tau}p_1,e^{-\tau}p_2,e^{-\tau}p_3)G(\tau;p_3)
\notag\\
   &\qquad{}
   -iT^aZ_\tau e^{-p_1^2+p_2^2-p_3^2}G(\tau;-p_1)
   \frac{1}{\Slash{p}_1}
   \left[
   \widetilde{C}(e^{-\tau}p_1)\Slash{p}_3
   +\widetilde{C}(e^{-\tau}p_3)\Slash{p}_1
   \right]
   \frac{1}{\Slash{p}_3}G(\tau;p_3)
   \frac{p_{2\mu}}{p_2^2}
\notag\\
   &\qquad{}
   -4iT^ae^{p_1^2+p_2^2-p_3^2}W(p_3;p_2,p_1)
   [1+G(\tau;-p_1)]G(\tau;p_3)p_{1\mu}
\notag\\
   &\qquad{}
   +4iT^ae^{-p_1^2+p_2^2+p_3^2}W(p_1;p_2,p_3)
   G(\tau;-p_1)[1+G(\tau;p_3)]p_{3\mu}
\notag\\
   &\qquad{}
   -4iT^ae^{p_1^2+p_2^2-p_3^2}W(p_3;p_2,p_1)
   [1+G(\tau;-p_1)]T(\tau;p_2)G(\tau;p_3)
\notag\\
   &\qquad\qquad{}
   \times
   \left(p_2^2\delta_{\mu\nu}-p_{2\mu}p_{2\nu}\right)p_{1\nu}
\notag\\
   &\qquad{}
   +4iT^ae^{-p_1^2+p_2^2+p_3^2}W(p_1;p_2,p_3)
   G(\tau;-p_1)T(\tau;p_2)[1+G(\tau;p_3)]
\notag\\
   &\qquad\qquad{}
   \times
   \left(p_2^2\delta_{\mu\nu}-p_{2\mu}p_{2\nu}\right)p_{3\nu},
\label{eq:(3.34)}
\end{align}
where we have used
\begin{equation}
   e^{\tau(4-D)}g_\tau^{-2}e^{-2p^2}T(\tau;p)C(e^{-\tau}p)=-1-p^2T(\tau;p),
\label{eq:(3.35)}
\end{equation}
which follows from~Eq.~\eqref{eq:(3.6)}, and~Eq.~\eqref{eq:(3.32)}.

\subsubsection{Locality}
Finally, we impose the locality on the Wilson action and determine the so far
arbitrary~$g_\mu^a$ in~Eq.~\eqref{eq:(3.34)}. For locality, the
function~$H_\mu^a$ in~Eq.~\eqref{eq:(3.34)} should be analytic at~$p_i=0$.
Since the functions~$T$ and~$G$ in~Eq.~\eqref{eq:(3.6)} and $W$
in~Eq.~\eqref{eq:(3.21)} are analytic, the term in~Eq.~\eqref{eq:(3.34)} that
is non-analytic is
\begin{equation}
   -iT^aZ_\tau e^{-p_1^2+p_2^2-p_3^2}G(\tau;-p_1)
   \frac{1}{\Slash{p}_1}
   \left[
   \widetilde{C}(e^{-\tau}p_1)\Slash{p}_3
   +\widetilde{C}(e^{-\tau}p_3)\Slash{p}_1
   \right]
   \frac{1}{\Slash{p}_3}G(\tau;p_3)\frac{p_{2\mu}}{p_2^2}.
\label{eq:(3.36)}
\end{equation}
Since $G(\tau;-p_1)\propto\Slash{p}_1$ and~$G(\tau;p_3)\propto\Slash{p}_3$, as
Eq.~\eqref{eq:(3.6)} shows, the only non-analyticity arises from the
factor~$1/p_2^2$. We have to choose the as yet undetermined function $g_\mu^a$
so that this singularity is cancelled.

Although it turns out that it is always possible to choose $g_\mu^a$ so that
the $1/p_2^2$ singularity is cancelled, the expression of such a $g_\mu^a$ for
the general case is very complicated and not illuminating. Here, therefore, we
are content with the expression for a particular case, i.e.\ the
limit~$\tau\to\infty$. In this limit, all irrelevant operators die out and the
expressions become much simpler. For~$\tau\to\infty$, because of the locality
of~$T$ and~$G$, $C(e^{-\tau}p)\to C_0$,
$\widetilde{C}(e^{-\tau}p)\to\widetilde{C}_0$, and
\begin{equation}
   T(\tau=\infty;p)\equiv T(p)
   =-\frac{1}{zC_0e^{-2p^2}+p^2},\qquad
   G(\tau=\infty;p)\equiv G(p)
   =-\frac{\Slash{p}}{Z\widetilde{C}_0e^{-2p^2}+\Slash{p}},
\label{eq:(3.37)}
\end{equation}
where
\begin{equation}
   z\equiv\lim_{\tau\to\infty}e^{\tau(4-D)}g_\tau^{-2},\qquad
   Z\equiv Z_{\tau=\infty}.
\label{eq:(3.38)}
\end{equation}
Equation~\eqref{eq:(3.36)} in this limit then becomes
\begin{equation}
   iT^aZ\widetilde{C}_0
   e^{-p_1^2+p_2^2-p_3^2}G(-p_1)\frac{1}{\Slash{p}_1}
   \gamma_\nu\frac{1}{\Slash{p}_3}G(p_3)
   \frac{p_{2\mu}p_{2\nu}}{p_2^2},
\label{eq:(3.39)}
\end{equation}
under the momentum conservation~$p_1+p_3=-p_2$. Then, the choice (recall that
$g_\mu^a$ must be transverse)
\begin{equation}
   e^{-2\tau}g_\mu^a(e^{-\tau}p_1,e^{-\tau}p_2,e^{-\tau}p_3)
   =-iT^aC_0\widetilde{C}_0\frac{1}{\Slash{p}_1}
   \gamma_\nu\frac{1}{\Slash{p}_3}
   \left(\delta_{\mu\nu}-\frac{p_{2\mu}p_{2\nu}}{p_2^2}\right)
\label{eq:(3.40)}
\end{equation}
cancels the non-analyticity in~Eq.~\eqref{eq:(3.34)} in the
limit~$\tau\to\infty$. In fact, with this choice,
\begin{align}
   &H_\mu^a(\tau=\infty;p_1,p_2,p_3)\equiv H_\mu^a(p_1,p_2,p_3)
\notag\\
   &=-iT^aZ\widetilde{C}_0
   e^{-p_1^2+p_2^2-p_3^2}
   G(-p_1)\frac{1}{\Slash{p}_1}
   \gamma_\nu\frac{1}{\Slash{p}_3}G(p_3)
   zC_0e^{-2p_2^2}T(p_2)
   \left(\delta_{\mu\nu}-\frac{p_{2\mu}p_{2\nu}}{p_2^2}\right)
\notag\\
   &\qquad{}
   +iT^aZ\widetilde{C}_0
   e^{-p_1^2+p_2^2-p_3^2}G(-p_1)\frac{1}{\Slash{p}_1}
   \gamma_\nu\frac{1}{\Slash{p}_3}G(p_3)
   \frac{p_{2\mu}p_{2\nu}}{p_2^2}
\notag\\
   &\qquad{}
   -4iT^ae^{p_1^2+p_2^2-p_3^2}W(p_3;p_2,p_1)
   [1+G(-p_1)]G(p_3)p_{1\mu}
\notag\\
   &\qquad{}
   +4iT^ae^{-p_1^2+p_2^2+p_3^2}W(p_1;p_2,p_3)
   G(-p_1)[1+G(p_3)]p_{3\mu}
\notag\\
   &\qquad{}
   -4iT^ae^{p_1^2+p_2^2-p_3^2}W(p_3;p_2,p_1)
   [1+G(-p_1)]G(p_3)
   T(p_2)\left(p_2^2\delta_{\mu\nu}-p_{2\mu}p_{2\nu}\right)p_{1\nu}
\notag\\
   &\qquad{}
   +4iT^ae^{-p_1^2+p_2^2+p_3^2}W(p_1;p_2,p_3)
   G(-p_1)[1+G(p_3)]
   T(p_2)\left(p_2^2\delta_{\mu\nu}-p_{2\mu}p_{2\nu}\right)p_{3\nu}.
\label{eq:(3.41)}
\end{align}
Since $zC_0e^{-2p_2^2}T(p_2)\to-1$ for~$p_2^2\to0$, the singularity~$1/p_2^2$ is
cancelled. Equation~\eqref{eq:(3.41)} provides the gauge-invariant local
expression of the interaction vertex~$H_\mu^a$ at~$\tau\to\infty$.\footnote{In
the low-momentum limit~$p_i\to0$, $H_\mu^a$ reduces to the conventional
expression of the vertex, $T^a/(iZ\widetilde{C}_0)\gamma_\mu$.} The resulting
Wilson action in~Eq.~\eqref{eq:(3.3)} depends on the parameters
$Z\widetilde{C}_0$, $zC_0$, and~$g_\tau=e^{\tau(4-D)}z^{-1/2}$. In the present
order of approximation, the first two are marginal and the last one is marginal
for~$D=4$ and relevant for~$D<4$.

Interestingly, if we assume that the function~$G$ satisfies the GW relation
in~$O(g_\tau^0)$, Eq.~\eqref{eq:(3.7)}, then the solution
in~Eq.~\eqref{eq:(3.41)} also satisfies the GW relation
in~Eq.~\eqref{eq:(2.22)} in~$O(g_\tau^1)$, i.e.
\begin{equation}
   \left[1+2G(-p_1)\right]\gamma_5H_\mu^a(p_1,p_2,p_3)
   +H_\mu^a(p_1,p_2,p_3)\gamma_5\left[1+2G(p_3)\right]=0.
\label{eq:(3.42)}
\end{equation}
This can be readily seen\footnote{We assume that Dirac matrices and~$\gamma_5$
anti-commute.} by noting Eq.~\eqref{eq:(3.32)} and
\begin{align}
   \left[1+2G(-p_1)\right]\gamma_5G(-p_1)&=G(-p_1)(-\gamma_5),
\notag\\
   G(p_3)\gamma_5\left[1+2G(p_3)\right]&=(-\gamma_5)G(p_3),
\label{eq:(3.43)}
\end{align}
which follow from~Eq.~\eqref{eq:(3.7)}. We do not have any understanding on
whether this is accidental or inevitable. It would be troublesome, however, if
Eq.~\eqref{eq:(3.41)} cannot fulfill the GW relation, because
Eq.~\eqref{eq:(3.41)} provides essentially the unique\footnote{One may
generalize the above gauge-invariant local solution by adding a local function
of~$e^{-\tau}p_i$ that is proportional
to~$e^{-2\tau}(p_2^2\delta_{\mu\nu}-p_{2\mu}p_{2\nu})$
to~$g_\mu^a(e^{-\tau}p_1,e^{-\tau}p_2,e^{-\tau}p_3)$ in~Eq.~\eqref{eq:(3.34)}.
Such a generalization, however, introduces only irrelevant operators to the
Wilson action and does not change $H_\mu^a(\tau=\infty;p_1,p_2,p_3)$.}
gauge-invariant local solution of the GFERG equation to~$O(g_\tau^1)$.

As demonstrated in~Appendix~\ref{sec:B}, we can obtain the axial anomaly
in~$D=2$ from the expression of the gauge-invariant local Wilson action
to~$O(A)$ given by~Eq.~\eqref{eq:(3.3)} with~Eq.~\eqref{eq:(3.41)}.

\section{Conclusion}
\label{sec:4}
We have formulated an ERG equation (the GFERG equation) in vector-like gauge
theories, Eq.~\eqref{eq:(2.23)}, on the basis of the notion of the gradient
flow and the fermion flow. The GFERG equation preserves the gauge invariance
and the chiral symmetry in a modified (\`a la Ginsparg--Wilson) form. The
formulation awaits applications such as the search for nontrivial fixed points
in gauge theory on the basis of a certain gauge-invariant truncation of the
Wilson action. Before going into such a nonperturbative study, however, we
should better understand perturbative aspects of the GFERG equation. In this
paper we obtained a gauge-invariant local Wilson action in QED by solving the
GFERG equation to~$O(g_\tau^1)$ at the Gaussian fixed point. We should pursue
this perturbative analysis at least to~$O(g_\tau^2)$, where we should be able
to observe ``quantum corrections.'' Another important remaining issue is the
finiteness of correlation functions in the continuum limit; we want to
understand this question in GFERG, in a manner similar to the argument
in~Ref.~\cite{Sonoda:2019ibh}, using the results
of~Refs.~\cite{Luscher:2011bx,Luscher:2013cpa} as a clue.

\section*{Acknowledgments}
We would like to thank Hidenori Sonoda for insightful remarks.
This work was partially supported by Japan Society for the Promotion of Science
(JSPS) Grant-in-Aid for Scientific Research Grant Number JP20H01903.

\appendix

\section{Manifestly chiral-invariant formulation}
\label{sec:A}
The GFERG formulation that is consistent with the conventional form of the
chiral transformation in~Eq.~\eqref{eq:(2.11)} can be obtained by setting
\begin{align}
   &e^{S_\tau[A,\psi,\Bar{\psi}]}
\notag\\
   &=
   \exp\left[
   \int d^Dx\,
   \frac{1}{2}
   \frac{\delta^2}{\delta A_\mu^a(x)\delta A_\mu^a(x)}\right]
   \exp\left[-i\int d^Dx'\,
   \frac{\delta}{\delta\psi(x')}\Slash{D}_\tau
   \frac{\delta}{\delta\Bar{\psi}(x')}
   \right]
\notag\\
   &\qquad{}
   \times
   \int[dA'd\psi'd\Bar{\psi}']\,
   \prod_{x'',\nu,b}\delta
   \left(A_\nu^b(x'')-e^\tau
   g_\tau^{-1}B_\nu^{\prime b}(t,x''e^\tau)\right)
\notag\\
   &\qquad\qquad{}
   \times
   \delta
   \left(
   \psi(x'')-e^{\tau(D-1)/2}Z_\tau^{1/2}\chi'(t,x''e^\tau)
   \right)
   \delta
   \left(
   \Bar{\psi}(x'')-e^{\tau(D-1)/2}Z_\tau^{1/2}\Bar{\chi}'(t,x''e^\tau)
   \right)
\notag\\
   &\qquad\qquad{}
   \times
   \exp\left[i\int d^Dx'''\,
   \frac{\delta}{\delta\psi'(x''')}\Slash{D}_{\tau=0}
   \frac{\delta}{\delta\Bar{\psi}'(x''')}
   \right]
   \exp\left[
   -\int d^Dx''''\,
   \frac{1}{2}
   \frac{\delta^2}
   {\delta A_\rho^{\prime c}(x'''')\delta A_\rho^{\prime c}(x'''')}\right]
\notag\\
   &\qquad\qquad{}
   \times
   \,e^{S_{\tau=0}[A',\psi',\Bar{\psi}']},
\label{eq:(A1)}
\end{align}
where $\Slash{D}_\tau$ is given by~Eq.~\eqref{eq:(2.15)}.

Taking the $\tau$~derivative of this, we have the following GFERG equation:
\begin{align}
   &\frac{\partial}{\partial\tau}
   e^{S_\tau[A,\psi,\Bar{\psi}]}
\notag\\
   &=\int d^Dx\,
   \frac{\delta}{g_\tau\delta A_\mu^a(x)}
\notag\\
   &\qquad{}
   \times
   \left.\left[
   -2D_\nu F_{\nu\mu}^a(x)-2\alpha_0D_\mu\partial_\nu A_\nu^a(x)
   -\left(\frac{D-2}{2}
   +\frac{\zeta_\tau}{2}
   +x\cdot\frac{\partial}{\partial x}\right)A_\mu^a(x)
   \right]\right|_{A\to g_\tau(A+\delta/\delta A)}
\notag\\
   &\qquad\qquad{}
   \times
   e^{S_\tau[A,\psi,\Bar{\psi}]}
\notag\\
   &\qquad{}
   +e^{S_\tau[A,\psi,\Bar{\psi}]}
\notag\\
   &\qquad{}
   \times
   \int d^Dx\,
   \tr\Biggl\{
   \left.\left[
   2\Delta-2\alpha_0\partial_\mu A_\mu(x)
   +\left(\frac{D-1}{2}+\frac{\eta_\tau}{2}+x\cdot\frac{\partial}{\partial x}
   \right)
   \right]\right|_{A\to g_\tau(A+\overleftarrow{\delta}/\delta A)}\psi(x)
\notag\\
   &\qquad\qquad\qquad\qquad{}
   \times\frac{\overleftarrow{\delta}}{\delta\psi(x)}
   \Biggr\}
\notag\\
   &\qquad{}
   +\int d^Dx\,
   \tr\left\{
   \frac{\delta}{\delta\Bar{\psi}(x)}\cdot\Bar{\psi}(x)
   \left.\left[
   2\overleftarrow{\Delta}
   +2\alpha_0\partial_\mu A_\mu(x)
   +\left(\frac{D-1}{2}+\frac{\eta_\tau}{2}
   +\frac{\overleftarrow{\partial}}{\partial x}\cdot x\right)
   \right]\right|_{A\to g_\tau(A+\delta/\delta A)}
   \right\}
\notag\\
   &\qquad\qquad\qquad\qquad{}
   \times
   e^{S_\tau[A,\psi,\Bar{\psi}]}
\notag\\
   &\qquad{}
   +\int d^Dx\,
   \tr\Biggl\{\left.\left[
   2\Delta-2\alpha_0\partial_\mu A_\mu(x)
   +\left(\frac{D-1}{2}+\frac{\eta_\tau}{2}
   +x\cdot\frac{\partial}{\partial x}\right)
   \right]i\Slash{D}
   \right|_{A\to g_\tau(A+\delta/\delta A)}\frac{\delta}{\delta\Bar{\psi}(x)}
\notag\\
   &\qquad\qquad\qquad\qquad{}
   \times
   e^{S_\tau[A,\psi,\Bar{\psi}]}
   \frac{\overleftarrow{\delta}}{\delta\psi(x)}
   \Biggr\}
\notag\\
   &\qquad{}
   +\int d^Dx\,
   \tr\Biggl\{
   \frac{\delta}{\delta\Bar{\psi}(x)}e^{S_\tau[A,\psi,\Bar{\psi}]}
   \frac{\overleftarrow{\delta}}{\delta\psi(x)}
\notag\\
   &\qquad\qquad\qquad\qquad{}
   \times
   \left.i\overleftarrow{\Slash{D}}
   \left[
   -2\overleftarrow{\Delta}
   -2\alpha_0\partial_\mu A_\mu(x)
   -\left(\frac{D-1}{2}+\frac{\eta_\tau}{2}
   +\frac{\overleftarrow{\partial}}{\partial x}\cdot x\right)
   \right]\right|_{A\to g_\tau(A+\overleftarrow{\delta}/\delta A)}
   \Biggr\}
\notag\\
   &\qquad{}
   +\int d^Dx\,
   \tr\Biggl\{
   i\gamma_\mu T^a
   \frac{\delta}{\delta\Bar{\psi}(x)}
   e^{S_\tau[A,\psi,\Bar{\psi}]}
   \frac{\overleftarrow{\delta}}{\delta\psi(x)}
\notag\\
   &\qquad\qquad\qquad\qquad{}
   \times   
   \left.\left[
   -2D_\nu F_{\nu\mu}^a(x)-2\alpha_0D_\mu\partial_\nu A_\nu^a(x)
   -\left(1
   +x\cdot\frac{\partial}{\partial x}\right)A_\mu^a(x)
   \right]\right|_{A\to g_\tau(A+\overleftarrow{\delta}/\delta A)}
   \Biggr\}.
\label{eq:(A2)}
\end{align}

\section{Axial anomaly in~$D=2$}
\label{sec:B}
We assume that the Wilson action is quadratic in the fermion field and set
\begin{equation}
   S_\tau=-\int d^Dx\,d^Dy\,\Bar{\psi}(x)D(x,y)\psi(y)+\dotsb.
\label{eq:(B1)}
\end{equation}
Then, from~Eq.~\eqref{eq:(2.21)}, the chiral invariance
$e^{-S_\tau}\Hat{\mit\Gamma}_5e^{S_\tau}=0$ implies the GW relation,
\begin{equation}
   \gamma_5D(x,y)+D(x,y)\gamma_5
   +2\int d^Dz\,D(x,z)\gamma_5D(z,y)=0.
\label{eq:(B2)}
\end{equation}
Under the infinitesimal chiral transformation in~Eq.~\eqref{eq:(2.11)} with the
localized parameter~$\alpha\to\alpha(x)$, the Wilson action
in~Eq.~\eqref{eq:(B1)} changes as
\begin{align}
   S_\tau
   &\to S_\tau
   -i\int d^Dx\,d^Dy\,
   [\alpha(y)-\alpha(x)]\Bar{\psi}(x)D(x,y)\gamma_5\psi(x)
\notag\\
   &\qquad{}
   +2i\int d^Dx\,d^Dy\,d^Dz\,
   \alpha(x)\Bar{\psi}(x)D(x,z)\gamma_5D(z,y)\psi(x),
\label{eq:(B3)}
\end{align}
where we have used the GW relation in~Eq.~\eqref{eq:(B2)}. Therefore, if the
integration measure $[d\psi d\Bar{\psi}]$ is invariant under the chiral
transformation, we have the identity
\begin{align}
   &\int d^Dx\,d^Dy\,[\alpha(y)-\alpha(x)]
   \left\langle
   \Bar{\psi}(x)D(x,y)\gamma_5\psi(y)
   \right\rangle_{S_\tau}
\notag\\
   &=2\int d^Dx\,d^Dy\,d^Dz\,\alpha(x)
   \left\langle
   \Bar{\psi}(x)D(x,z)\gamma_5D(z,y)\psi(y)
   \right\rangle_{S_\tau},
\label{eq:(B4)}
\end{align}
and for a \emph{fixed gauge-field configuration}, the right-hand side of this
expression is computed as
\begin{align}
   &2\int d^Dx\,d^Dy\,d^Dz\,\alpha(x)
   \left\langle
   \Bar{\psi}(x)D(x,z)\gamma_5D(z,y)\psi(y)
   \right\rangle_{S_\tau}
\notag\\
   &=-2\int d^Dx\,d^Dy\,d^Dz\,\alpha(x)
   \tr\left[D(x,z)\gamma_5D(z,y)\left\langle
   \psi(y)\Bar{\psi}(x)
   \right\rangle_{S_\tau}\right]
\notag\\
   &=-2\int d^Dx\,\alpha(x)
   \tr\left[\gamma_5 D(x,x)\right],
\label{eq:(B5)}
\end{align}
where we have used $\int d^Dy\,D(z,y)\langle\psi(y)\Bar{\psi}(x)\rangle_{S_\tau}
=\delta^{(D)}(z-x)$.

Now, for the parametrization of the Wilson action in~Eq.~\eqref{eq:(3.3)}, we
find
\begin{equation}
   D(x,x)
   =\int_\ell G(\tau;\ell)
   -g_\tau\int_pe^{ipx}A_\mu^a(p)\int_\ell
   H_\mu^a(\tau;-\ell-p,p,\ell)+\dotsb
\label{eq:(B6)}
\end{equation}
and thus the factor in~Eq.~\eqref{eq:(B5)} is given by
\begin{equation}
   \tr\left[\gamma_5 D(x,x)\right]
   =\tr\left\{
   \gamma_5
   \left[\int_\ell G(\tau;\ell)
   -g_\tau\int_pe^{ipx}A_\mu^a(p)\int_\ell
   H_\mu^a(\tau;-\ell-p,p,\ell)+\dotsb\right]\right\}.
\label{eq:(B7)}
\end{equation}
The term containing~$H_\mu^a$, being linear in the gauge potential, is relevant
to the axial anomaly in $D=2$.

So far, all elements in this paper have been dimensionless, i.e.\ everything is
measured in units of a UV cutoff~$\Lambda_0$. What we are eventually interested
in is the continuum limit~$\Lambda_0\to\infty$, in which the external momentum
carried by the gauge potential in physical units, $p\Lambda_0$, is kept fixed.
The axial anomaly in this ``classical continuum limit'' is given by a
low-momentum limit of~Eq.~\eqref{eq:(B7)}. Then, in~Eq.~\eqref{eq:(B7)}, noting
that $\tr(\gamma_5\Slash{\ell}^n)=0$, we see that one can replace $H_\mu^a$
given in~Eq.~\eqref{eq:(3.41)} (we consider the Wilson action
with~$\tau\to\infty$ that is relevant in the continuum limit) by
\begin{align}
   &H_\mu^a(\tau=\infty;-\ell-p,p,\ell)
\notag\\
   &=-iT^aZ\widetilde{C}_0e^{-(\ell+p)^2+p^2-\ell^2}
   G(\ell+p)\frac{1}{\Slash{\ell}+\Slash{p}}
   \gamma_\mu\frac{1}{\Slash{\ell}}G(\ell)
\notag\\
   &\qquad{}
   +2iT^a\left[1+G(\ell+p)\right]G(\ell)\ell_\mu
   +2iT^aG(\ell+p)\left[1+G(\ell)\right]\ell_\mu+O(p^2),
\label{eq:(B8)}
\end{align}
where we have noted that $\lim_{p\to0}W(\ell;p,-\ell-p)=
\lim_{p\to0}W(-\ell-p;p,\ell)=1/2$. Using this, after taking the trace over
Dirac indices by~$\tr(\gamma_5\gamma_\mu\gamma_\nu)=
2i\epsilon_{\mu\nu}$ for~$D=2$ (we set $\gamma_5\equiv-i\gamma_0\gamma_1$
and~$\epsilon_{01}=1$), we have\footnote{Here, we are assuming an Abelian
gauge theory, for which the gauge-group generator is given by~$T^a=-ie$
as~Eq.~\eqref{eq:(3.1)}.}
\begin{align}
   &\int_\ell\tr\left[\gamma_5H_\mu^a(\tau=\infty;-\ell-p,p,\ell)\right]
\notag\\
   &=2T^aZ^2\widetilde{C}_0^2
   \int_\ell
   \frac{e^{-4\ell^2}(1+4\ell^2)}
   {(-Z^2\widetilde{C}_0^2e^{-4\ell^2}+\ell^2)^2}\epsilon_{\mu\nu}p_\nu
   +O(p^2)
\notag\\
   &=-\frac{1}{2\pi}
   T^a\mathcal{I}(-Z^2\widetilde{C}_0^2)\epsilon_{\mu\nu}p_\nu
   +O(p^2),
\label{eq:(B9)}
\end{align}
where the integral
\begin{equation}
   \mathcal{I}(\xi)\equiv
   \int_0^\infty dx\,
   \frac{\xi e^{-4x}(1+4x)}{(\xi e^{-4x}+x)^2}
   =-\int_0^\infty dx\,\frac{d}{dx}
   \frac{\xi}{e^{4x}x+\xi}
   =1
\label{eq:(B10)}
\end{equation}
is independent of~$\xi>0$.

When the external momentum is small, on the other hand, one sees that the
left-hand side of~Eq.~\eqref{eq:(B4)} reduces to
\begin{equation}
   \frac{1}{iZ\widetilde{C_0}}
   \int d^Dx\,\alpha(x)
   \partial_\mu
   \left\langle\Bar{\psi}(x)\gamma_\mu\gamma_5\psi(x)\right\rangle_{S_{\tau=\infty}}
   \equiv
   \int d^Dx\,\alpha(x)
   \partial_\mu
   \left\langle j_{5\mu}(x)\right\rangle_{S_{\tau=\infty}}.
\label{eq:(B11)}
\end{equation}
Since the action is normalized
as~$S_{\tau=\infty}=
1/(iZ\widetilde{C}_0)\int d^Dx\,\Bar{\psi}(x)\Slash{\partial}\psi(x)+\dotsb$
in the low-momentum limit, Eq.~\eqref{eq:(B11)} gives the total divergence of a
\emph{correctly normalized\/} axial-vector current~$j_{5\mu}(x)$. Thus,
finally, combining Eqs.~\eqref{eq:(B11)}, \eqref{eq:(B4)}, \eqref{eq:(B5)},
\eqref{eq:(B7)}, and~\eqref{eq:(B9)}, we have the axial anomaly in~$D=2$ as
\begin{equation}
   \partial_\mu
   \left\langle j_{5\mu}(x)\right\rangle_{S_{\tau=\infty}}
   =-\frac{i}{\pi}g_{\tau=\infty}T^a\epsilon_{\mu\nu}\partial_\mu A_\nu^a(x).
\label{eq:(B12)}
\end{equation}
As anticipated from the gauge invariance and locality of our Wilson action,
this reproduces the correct expression of the axial anomaly in~$D=2$.

\end{document}